\pdfoutput=1

\documentclass[a4paper]{article}

\usepackage{amsmath}
\usepackage{amssymb}
\usepackage{amstext}
\usepackage{mathtools}
\usepackage{graphicx}
\usepackage{fullpage}
\usepackage{array}
\usepackage{tabularx}
\usepackage{multirow}
\usepackage{longtable}
\usepackage{xcolor}
\DeclareMathOperator{\sign}{sgn}

\usepackage{float}
\usepackage[normalem]{ulem}

\usepackage{textcomp}%
\usepackage{manyfoot}%
\usepackage{booktabs}%
\usepackage{algorithm}%
\usepackage{algorithmicx}%
\usepackage{algpseudocode}%
\usepackage{authblk}

\makeatletter
\newcommand{\ALOOP}[1]{\ALC@it\algorithmicloop\ #1%
  \begin{ALC@loop}}
\newcommand{\ENDALOOP}{\end{ALC@loop}\ALC@it\algorithmicendloop}

\newcommand{\algorithmicbreak}{\textbf{break}}

\makeatother
\usepackage{listings}%

\newcommand{\bs}[1]{\boldsymbol{#1}}

\def\bsf{\bs{f}}

\def\bfU{\mathbf{U}}
\def\bfB{\mathbf{B}}
\def\bfC{\mathbf{C}}
\def\bfb{\mathbf{b}}
\providecommand{\keywords}[1]
{
  \small	
  \textbf{\textit{Keywords--}} #1
}

\begin{document}


\title{Isogeometric Analysis for the Pricing of Financial Derivatives with Nonlinear Models: Convertible Bonds and Options}

\author[1,2]{Rakhymzhan Kazbek}
\author[3]{Yogi Erlangga}
\author[2]{Yerlan Amanbek}
\author[2]{Dongming Wei}

\affil[1]{Astana IT University, Department of Computation and Data Science, Mangilik El C1, Astana, Kazakhstan}
\affil[2]{Department of Mathematics, School of Sciences and Humanities, Nazarbayev University, Kabanbay Batyr 53, Astana, Kazakhstan}
\affil[3]{Zayed University, Department of Mathematics, Abu Dhabi Campus, P.O. Box 144534, United Arab Emirates}

\date{} 

\maketitle

\begin{abstract}
Computational efficiency is essential for enhancing the accuracy and practicality of pricing complex financial derivatives. In this paper, we discuss Isogeometric Analysis (IGA) for valuing financial derivatives, modeled by two nonlinear Black-Scholes PDEs: the Leland model for European call with transaction costs and the AFV model for convertible bonds with default options. We compare the solutions of IGA with finite difference methods (FDM) and finite element methods (FEM). In particular, very accurate solutions can be numerically calculated on far less mesh (knots) than FDM or FEM, by using non-uniform knots and weighted cubic NURBS, which in turn reduces the computational time significantly.
\end{abstract}

\section{Introduction} 
\label{sec:introduction}
Isogeometric analysis (IGA) is a finite-element method (FEM) that utilizes a special type of basis functions, known as non-uniform rational B-splines or NURBS~\cite{DeBoor}. First introduced by Hughes in~\cite{Hughes2005, IGA_book}, IGA has since gained popularity as a robust FEM approach. Among  its key strengths are the ability to deal with complex geometries and the natural link it creates between FEM and computer-aided design (CAD), a tool widely used in geometric design and mesh generation. Moreover, IGA’s use of continuous and sufficiently differentiable NURBS basis functions enables it to produce higher-order smooth solutions within the computational domain  $\Omega$. In contrast, traditional FEM methods are limited in this regard, as they employ piecewise continuous basis functions. For instance, the linear P1-FEM solution achieves only  $C^0$-continuous across the domain $\Omega$. In problems where derivatives of the solution are critical, such as in financial hedging strategies involving option prices (the Greeks), having only a $C^0(\Omega)$-continuous solution makes the calculation of derivatives conceptually not straight-forward. These derivatives can only be determined within the interior of elements, rendering their values non-existent at the boundaries.

In option pricing, the finite difference method (FDM) is widely favored by financial practitioners due to its computational efficiency and straightforward implementation~\cite{practitioners,Christara2022,Zhu_bonds_ADI}. Moreover, FDM is well-suited to addressing non-linearities in the partial differential equations (PDEs) associated with option pricing. While higher-order FDMs can potentially enhance solution accuracy with fewer grid points, they often involve extended discretization stencils, which can complicate the treatment of boundary conditions. On the other hand, finite element methods (FEM) provide a more natural framework for achieving higher-order approximations and managing boundary conditions effectively. Despite these advantages, FEM has not gained the same level of popularity as FDM in option pricing~\cite{Highly_accurate_fem,Wei2024,kazbek2023valuation}. Although FEM typically offers superior convergence rates compared to FDM~\cite{realoption_FDM_FEM}, the use of Lagrange polynomial basis functions in standard FEM formulations can result in highly oscillatory solutions.

In contrast to the Lagrange polynomial basis functions used in the standard FEM, IGA utilizes NURBS, a class of functions that satisfy certain differentiability conditions across elements, as the basis functions. For instance, cubic NURBS basis functions are twice differentiable in the computational domain $\Omega$. Furthermore, the NURBS basis functions are nonnegative and non-oscillatory. By using proper weighting, NURBS basis functions can very accurately approximate (or fit) non-smooth functions, such as the pay-off function in option pricing, a feature not found in FDM or standard FEM. While the success of IGA has been demonstrated in a wide range of applications~\cite{IGA_for_jump_diff, IGA_biology,IGA_nanomedicine,IGA_KirchhoffLoveshell,IGA_bioenginerring}, there is not so much work on using IGA for (nonlinear) option pricing. We should in this regard mention the work of~\cite{quintic_Bspline}, which uses a special case of NURBS, called B-Splines, to solve the linear Black-Scholes equation, and the recent work on IGA for European option~\cite{IGA_for_jump_diff}.

In this paper, we demonstrate IGA as an efficient alternative to the popular FDM and FEM for solving nonlinear partial differential equations (PDEs) arising in option pricing. In particular, with optimal weights and non-uniform knots, we show that NURBS IGA produces solutions that match accurately the solutions of FDM and FEM, but with low numbers of meshes/knots (hence, basis functions). The latter results in significant gain in computational time. For this purpose, out of many nonlinear models that are presently used for different types of options, we shall in particular consider two models: 
\begin{enumerate}
    \item[(i)] the Leland model~\cite{Leland}, which arises in European call option with transaction costs;
    \item[(ii)] the AFV model~\cite{Ayache2003}, used for convertible bond pricing with coupon payments. 
\end{enumerate}

The AFV model is an example of exotic options due to its variable parameters and inequality constraints, which influence the solution at each time step. Additionally, it exhibits path-dependency because of its embedded bond and equity option properties, where the price depends not only on the final payoff value but also on the entire price path computed at each time step. Moreover, the model allows early exercise (American options), which is represented by free boundaries. All of these lead to a complex structure from a computational perspective due to the changing payoff values in the inequality constraints. One approach for solving free boundaries such as in the AFV model is by reformulating the problems as Linear Complementarity Problems (LCPs)~\cite{Forsyth2002}, which can be solved using, e.g., ADI-based predictor-corrector schemes~\cite{Zhu_Landau} or through analytical approximations~\cite{kim1990analytic,broadie1996american,carr1992alternative,carr1998randomization}. Alternatively, one can formulate the model as an optimal control problem~\cite{Forsyth2007} using discrete Hamilton-Jacobi-Bellman (HJB) PDEs. The latter is often solved by using policy iteration algorithms~\cite{Christara2022}, which eliminate the need to explicitly determine the optimal exercise boundary at each time step. In this study, we employ formulation of the AFV model into a penalty PDE~\cite{Forsyth2002}, resulting in a system of nonlinear PDEs and transforming the free-boundary problems into ones defined on a fixed domain.

The paper is organized as follows. In Section~\ref{sec:methodology}, we discuss briefly the IGA method, followed by the construction of an IGA method for the nonlinear Leland PDE and the nonlinear AFV model. In Section~\ref{sec:timeintegral}, we discuss the time integration procedure, which is based on the $\theta$-scheme with linearization. Numerical results are presented and discussed in Section~\ref{sec:numresults}. We shall draw conclusions and outlook in  Section~\ref{sec:conclusion}. 

In the remaining part of this section, we briefly discuss the two PDEs used as the model problems in this paper.

\subsection{European contract with transaction costs: The Leland model}

The European call option price with transaction costs and with the strike price $\hat{K}$ at the terminal time $T$ can be modeled by the nonlinear PDE
\begin{equation}
    \displaystyle \frac{\partial V}{\partial t} + \frac{1}{2}\sigma^2S^2\left(1 + \textit{Le}\sign\left(\frac{\partial^2 V}{\partial S^2}\right)\right)\frac{\partial^2 V}{\partial S^2} +rS\frac{\partial V}{\partial S} - rV = 0, \label{eq:Leland}
\end{equation}
where $V = V(S,t)$ is the option price at any time $t \in (0,T)$, $S \ge 0$ is the value of the underlying asset, $r$ is risk-free interest rate, and  $\sigma$ is volatility. The constant 
$$
Le = \sqrt{\frac{2}{\pi}}\frac{c}{\sigma\sqrt{\delta t}}
$$
is called the Leland number, with $c$ be the round trip of transaction costs per currency limit. This model, introduced by Leland~\cite{Leland}, can be viewed as an adjustment to the standard linear Black-Scholes model, which assumes no transaction cost and continuous portfolio rebalancing. With transaction costs, such a continuous rebalancing becomes unrealistic in practice. The model~\eqref{eq:Leland} assumes transaction costs proportional to the asset value and allows a finite number of rebalancing at $\delta t$ towards maturity.

As in the linear Black-Scholes model, the PDE~\eqref{eq:Leland} is equipped with the following terminal and boundary conditions: 
\begin{align}
    V(S,T) &= \max(S-\hat{K},0), \label{eq:tcLeland} \\[5pt]
    V(0,t) &= 0 ~~\text{as}~S\rightarrow 0, \label{eq:bcLeland1}\\[5pt]
    V(S,t) &\approx S ~~\text{as}~S\rightarrow \infty. \label{eq:bcLeland2}
\end{align}


\subsection{American Convertible bond contract: the AFV model}

Convertible bonds subject to default strategies can be priced by the  so-called AFV model, developed in \cite{Ayache2003}. This model is considered an improvement from the earlier model, called the TF model, proposed in~\cite{vcb}, which was argued to be inconsistent with the derivative market requirements. The intuitive continuation from the TF to the AFV model involves the development of partial and total default strategies.

Following~\cite{Ayache2003}, the pricing of convertible bond subject to default strategies of the underlying asset is modeled by the system of PDEs:
\begin{align}
&\displaystyle \frac{\partial U}{\partial t}+ \frac{1}{2}\sigma^2S^2\frac{\partial^2 U}{\partial S^2} + (r+p\eta)S\frac{\partial U}{\partial S}-(r+p)U + p\max(kS(1-\eta),\hat{R}B) = 0,\label{Ueqxn}\\
&\displaystyle \frac{\partial B}{\partial t}+ \frac{1}{2}\sigma^2S^2\frac{\partial^2 B}{\partial S^2} + (r+p\eta)S\frac{\partial B}{\partial S}- (r+p)B + \hat{R}pB =0, \label{Veqxn}\\
&\displaystyle \frac{\partial C}{\partial t}+ \frac{1}{2}\sigma^2S^2\frac{\partial^2 C}{\partial S^2} + (r+p\eta)S\frac{\partial C}{\partial S}-(r+p)C + p\max(kS(1-\eta)- \hat{R}B,0) =0,\label{Ceqxn}
\end{align}
for the time $t\in (0,T)$ and the underlying stock price $S \ge 0$, where $U$ is the value of the convertible bond, $B$ is the bond component, and $C$ is the equity component. Furthermore, the parameters $r$, $p$, $\eta$, $\hat{R}$, $k$, and $\sigma$ are respectively the risk-free rate, the hazard rate, the indicator of the partial or total default strategies, the recovery factor upon the default,  the conversion rate, and  volatility. 

The terminal conditions at the maturity time $T$ are given as follows, with $F$ be the face value and $K_{coup}$ be the coupon rate,
\begin{eqnarray}
U(S,T) = 
\begin{cases}
 F+K_{coup}, &\text{if}\quad F+K_{coup}\geq kS,\\[5pt]
 kS, &\text{otherwise}, 
\end{cases} \label{TcondU} 
\end{eqnarray}
\begin{eqnarray}
B(S,T) = F+K_{coup},\label{TcondV} 
\end{eqnarray}
and
\begin{eqnarray}
C(S,T) = 
\begin{cases}
kS -F - K_{coup}, &\text{if}\quad kS -F - K_{coup}\geq 0,\\[5pt]
0,&\text{otherwise}.
\end{cases} \label{TcondC} 
\end{eqnarray}
Once the convertible bond is issued, the holder can convert it to assets before its expiration date, if it is reasonable, and the issuer will pay principal. Otherwise, the following inequalities hold throughout its lifetime:
\begin{enumerate}
\item Value of the convertible bond $U$ in \eqref{Ueqxn}, subject to the constraints:
\begin{eqnarray}
&U(S,t)\geq \max(B_{put},kS), \label{constraint:u_constrain_1}\\ [5pt]
&U(S,t)\leq \max(B_{call},kS).
\label{constraint:u_constrain_2}
\end{eqnarray}

\item  Value of the bond component $B$ in \eqref{Veqxn}, subject to the constraints:
\begin{eqnarray}
&B(S,t)\leq B_{call},\label{constraint:v_constrain_1}\\[5pt]
&B(S,t) + C(S,t)\geq B_{put}.
\label{constraint:v_constrain_2}
\end{eqnarray}

\item Value of the equity component $C$ in \eqref{Ceqxn}, subject to the constraints:
\begin{eqnarray}
&B(S,t) + C(S,t)\leq \max(B_{call},kS),\label{constraint:c_constrain_1}\\[5pt]
&B(S,t) + C(S,t)\geq kS.
\label{constraint:c_constrain_2}
\end{eqnarray}
\end{enumerate}

In~\eqref{constraint:u_constrain_2}--\eqref{constraint:c_constrain_2}, $B_{call}$ and $B_{put}$ are the dirty call and put price, respectively, which depend on the interest, accrued during the period of pending coupon payment. Under this scenario 
\begin{eqnarray}
   B_{put,call}(t) = B_{put,call}^{clean} + AccI(t),
\end{eqnarray}
where
\begin{eqnarray}
   AccI(t) = K_i \frac{t - t_{i-1}}{t_i - t_{i-1}}
\end{eqnarray}
is the accrued interest at any time $t$ between the time of the last coupon payment $t_{i-1}$ and the time of the next coupon payment $t_i$.

At $S = 0$, the boundary condition is given by the following system of time dependent equations, obtained by setting $S=0$ in~\eqref{Ueqxn}--\eqref{Ceqxn}:
\begin{eqnarray}
\begin{cases}
\displaystyle \frac{\partial U(0,t)}{\partial t} =  (r+p)U(0,t) - \hat{R}pB(0,t), \\[10pt]
\displaystyle \frac{\partial B(0,t)}{\partial t} = (r+(1-\hat{R})p) B(0,t),\\[10pt]
\displaystyle \frac{\partial C(0,t)}{\partial t} = (r+p) C(0,t).
\end{cases} \label{bcond0}
\end{eqnarray}
As $S\rightarrow\infty $, the boundary conditions represent the action of the investor, which will convert the bond to assets with the conversion ratio $k$: 
\begin{eqnarray}
\lim_{S \to \infty}
\begin{cases}
U(S,t)=kS, \\
B(S,t)=0,\\
C(S,t)=kS.
\end{cases} \label{bcond1}
\end{eqnarray}
 




\subsection{Transformation of the models}


\subsubsection{The Leland model}

Following the standard computational approach, we first transform the terminal-boundary value problem~\eqref{eq:Leland}--\eqref{eq:bcLeland2} to an initial-boundary value problem by applying the following change of variables~\cite{Wei2024}:
\begin{itemize}
  \item $\displaystyle \tau = \frac{1}{2} \sigma^2 (T - t)$;
  \item $\displaystyle x = \ln(S) + \kappa \tau $; and
  \item $\hat{v}(x,\tau) = e^{\kappa \tau} V(S,t)$,
\end{itemize}
where $\kappa = 2r/\sigma^2$. It can then be shown that the PDE~\eqref{eq:Leland} is transformed to 
\begin{equation}
    \displaystyle \frac{\partial \hat{v}}{\partial \tau} = \frac{\partial^2 \hat{v}}{\partial x^2} - \frac{\partial \hat{v}}{\partial x} + \textit{Le}\left|\frac{\partial^2 \hat{v}}{\partial x^2} - \frac{\partial \hat{v}}{\partial x}\right|= 0, \quad (\tau,x) \in (0,\frac{1}{2}\sigma^2 T) \times (-\infty,\infty),\label{eq:Lelandtrans}
\end{equation}
after using the fact that $\sign(\cdot) \times \cdot = |\cdot|$. For the construction of our IGA, we shall consider a mixed formulation of~\eqref{eq:Lelandtrans}, given by
\begin{eqnarray}
\begin{cases}
    \displaystyle \frac{\partial \hat{v}}{\partial \tau} = \frac{\partial^2 \hat{v}}{\partial x^2} - \frac{\partial \hat{v}}{\partial x} + \textit{Le}\left|\Tilde{v}\right|,\\[10pt]
     \displaystyle \Tilde{v} = \frac{\partial^2 \hat{v}}{\partial x^2} - \frac{\partial \hat{v}}{\partial x}.
\end{cases}
\label{eq:system_of_transformed_eqxns_for_Leland}
\end{eqnarray}
The terminal and the boundary conditions are transformed to the initial and boundary conditions
\begin{align}
    \hat{v}(x,0) &= \max(e^x-\hat{K},0),\\[5pt]
    \hat{v}(x,\tau) &= 0 ~~\text{as}~x\rightarrow -\infty,\\[5pt]
    \hat{v}(x,\tau) &\approx e^x - \hat{K} ~~\text{as}~x\rightarrow \infty.
\end{align}

\subsubsection{AFV model}

For the AFV model, we consider the change of variables~\cite{Zhu_Landau}:
\begin{itemize}
  \item $\displaystyle \tau = T - t $, and
  \item $\displaystyle x = \ln\left(\frac{S}{S_{\text{int}}}\right)$,
\end{itemize}
where $S_{\text{int}}$ is the stock price at the initial time $t = 0$. Applying the change of variables to the PDEs \eqref{Ueqxn}--\eqref{Ceqxn} results in the transformed PDEs
\begin{align}
\displaystyle\frac{\partial U}{\partial \tau} &= \frac{\sigma^2}{2}\frac{\partial^2 U}{\partial x^2} +\left(r+p\eta -\frac{\sigma^2}{2}\right)\frac{\partial U}{\partial x} - (r+p)U + p\delta, \label{eq:transU} \\[10pt]
\displaystyle\frac{\partial B}{\partial \tau} &= \frac{\sigma^2}{2}\frac{\partial^2 B}{\partial x^2} + \left(r+p\eta -\frac{\sigma^2}{2}\right)\frac{\partial B}{\partial x} - (r+p)B +\hat{R}pB,\label{eq:transB}\\[10pt]
\displaystyle\frac{\partial C}{\partial \tau} &= \frac{\sigma^2}{2}\frac{\partial^2 C}{\partial x^2} + \left(r+p\eta -\frac{\sigma^2}{2}\right)\frac{\partial C}{\partial x} - (r+p)C + p\gamma,\label{eq:transC}
\end{align}
where
\begin{align}
\delta = \begin{cases}
                   \hat{R}B,&\text{if } \hat{R}B > kS_{\text{int}}e^x(1-\eta), \\[5pt]
                   kS_{\text{int}}e^x(1-\eta), &\text{otherwise},
               \end{cases}
\end{align}
and
\begin{align}
    \displaystyle \gamma = \begin{cases}
                   0, &\text{if } kS_{\text{int}}e^x(1-\eta) - \hat{R}B < 0, \\[5pt]
                  kS_{\text{int}}e^x(1-\eta) - \hat{R}B , &\text{otherwise},
               \end{cases}
\end{align}
with $(\tau,x) \in (0,T)\times (-\infty,\infty)$.

The terminal, boundary conditions and inequality constraints can be transformed in the same manner.

To construct the penalty PDEs for the convertible bond valuation, we  reformulate the inequality constraint ~\eqref{constraint:u_constrain_1} as follows \cite{Forsyth2002}:
$$
   U(x,\tau) \ge \max(B_{put}, kS_{\text{int}}e^x) \, \Rightarrow \, \Pi_{put} :=  U_{put}^{\star} - U(x,\tau) \le 0,
$$
where $U_{put}^{\star} = \max(B_{put}, kS_{\text{int}}e^x)$. With~\eqref{constraint:u_constrain_2} reformulated as $\Pi_{call} := U(x,\tau) - U_{call}^{\star} \le 0$, where $U_{call}^{\star} = \max(B_{call},kS_{\text{int}}e^x)$, the PDE for the convertible bond valuation (Eq.~\eqref{eq:transU}) can be written as the penalty PDE
\begin{align}
    \frac{\partial U}{\partial \tau} &= \frac{\sigma^2}{2}\frac{\partial^2 U}{\partial x^2} +\left(r +p\eta - \frac{\sigma^2}{2}\right)\frac{\partial U}{\partial x}-(r+p)U + p\delta +\rho\max(U - U_{call}^{\star},0) + \rho\max(U_{put}^{\star} - U,0)  \notag\\     
    &= \frac{\sigma^2}{2}\frac{\partial^2 U}{\partial x^2}+\left(r +p\eta - \frac{\sigma^2}{2}\right)\frac{\partial U}{\partial x} - (r+p)U  + p\delta+ \rho \alpha_{call}\Pi_{call}  + \rho \alpha_{put} \Pi_{put}, \label{eq:transpenU}
\end{align}
where $\rho > 0$ is the penalty parameter,
\begin{eqnarray}
    \alpha_{call} = \begin{cases}
                   1,& \text{if } U-U_{call}^{\star} \ge 0, \\[5pt]
                   0,& \text{otherwise},
               \end{cases}  \text{ \, and \, }
    \alpha_{put} = \begin{cases}
                   1,& \text{if } U_{put} ^{\star} -U\ge 0, \\[5pt]
                   0,& \text{otherwise}.
               \end{cases}  \label{eq:alpha}
\end{eqnarray}
\section{Isogeometric analysis}
 \label{sec:methodology}

To describe IGA for solving the two nonlinear PDE models given by the system~\eqref{eq:system_of_transformed_eqxns_for_Leland} and the set of PDEs~\eqref{eq:transpenU},~\eqref{eq:transB}, and~\eqref{eq:transC}, we consider the unified PDE
\begin{align}
    \frac{\partial w}{\partial \tau } = \Upsilon_{1,w} \frac{\partial^2 w}{\partial x^2} + \Upsilon_{2,w} \frac{\partial w}{\partial x} - \Upsilon_{3,w} w + \mathcal{N}_w, \label{eq:unified}
\end{align}
where $w \in \{\hat{v},U,B,C\}$, with
\begin{align}
    \Upsilon_{1,w} =\begin{cases}
        1,& w = \hat{v}, \\
        \sigma^2/2,&\text{otherwise},
    \end{cases}
    \quad
    \Upsilon_{2,w} = \begin{cases}
       -1,& w = \hat{v}, \\
       r + p \eta - \sigma^2/2,& \text{otherwise},
    \end{cases} 
    \quad
    \Upsilon_{3,w} = \begin{cases}
       0,& w = \hat{v}, \\
       (r + p) - \hat{R} p, & w = B, \\
       r + p, &\text{otherwise},
    \end{cases} \notag
\end{align}
and
\begin{align}
\mathcal{N}_w = \begin{cases}
       Le|\tilde{v}|,& w = \hat{v}, \\
       p\delta + \rho(\alpha_{call} \Pi_{call} + \alpha_{put} \Pi_{put}), & w = U, \\
       0, &w = B, \\
       p \gamma, &w = C.
    \end{cases} \notag
\end{align}

\subsection{Weak formulation}


For $\Omega \subset \mathbb{R}$, let
\begin{align}
  H^1(\Omega) = \left\{ w: \Omega \to \mathbb{R} \, | \,w, \frac{\partial w}{\partial x} \in L^2(\Omega) \right\}. \notag
\end{align}
Associated with $H^1(\Omega)$ are the space of trial solutions $H^1_D(\Omega)$ and, respectively, the space of weighting functions  
$H^1_0(\Omega)$ defined as
$$
\displaystyle H^1_D (\Omega) = \left\{ w~|~ w\in H^1(\Omega),~ w~\text{satisfies Dirichlet condition on}~\partial\Omega\right\}$$
 and
$$
\displaystyle H^1_0(\Omega) = \left\{z_w~|~z_w\in H^1(\Omega),~ \displaystyle z_w|_{\partial\Omega} = 0 \right\}.
$$
The variational form of~\eqref{eq:unified} is obtained by multiplying it by an arbitrary test function $z_w\in H^1_0(\Omega)$ and integrating over $\Omega$, yielding
\begin{align}
    \int_{\Omega} \frac{\partial w }{\partial \tau } z_w = \int_{\Omega} \left( \Upsilon_{1,w} \frac{\partial^2 w}{\partial x^2} + \Upsilon_{2,w} \frac{\partial w}{\partial x} - \Upsilon_{3,w} w + \mathcal{N}_w \right) z_w, ~~~\forall z_w\in H^1_0(\Omega).
\end{align}
By applying the divergence theorem on the above equation, the pricing problem~\eqref{eq:unified} is now equivalent to finding $w\in H^1_D(\Omega)$ that satisfies the weak formulation
\begin{align}
    \frac{\partial}{\partial \tau} (w,z_w) = - \Upsilon_{1,w} (\nabla w, \nabla z_w) - \Upsilon_{2,w} (w, \nabla z_w) - \Upsilon_{3,w} (w,z_w) + (\mathcal{N}_w, z_w), \quad \forall z_w\in H^1_0(\Omega),
    \label{eq:Weak_form}
\end{align}
with $(f,g)$ denoted the $L^2(\Omega)$ inner product of two functions $f$ and $g$.

Let the solution $w$ be approximated by the linear combination 
\begin{align}\label{eq:lin.comb. of w}
   w \approx w_h = \sum_j w_j \phi_j =  \sum_{j=1}^{n} w_j \phi_j + \sum_{j \in \mathcal{I}_{\partial \Omega} } w_j \phi_j  \in S_D \subset H^1_D,
\end{align}
where the trial functions \{$\phi_j\}$, $j=1,\dots,n \in \mathcal{I}_{\Omega}$ form a basis for the $n$-dimensional space $S_0 \subset H^1_0$ and the second sum on the right-hand side interpolates the boundary data, with $\mathcal{I}_{\partial \Omega}$ denoting the set of indices of functions used to interpolate the boundary data. Setting $z_w = \phi_i$ (the Galerkin method),
the weak form~\eqref{eq:Weak_form} can be written as
\begin{align}
    \frac{\partial}{\partial \tau} \sum_j w_j (\phi_j,\phi_i) = - \Upsilon_{1,w} \sum_j w_j(\nabla \phi_j, \nabla \phi_i) - \Upsilon_{2,w} \sum_j w_j (\phi_j, \nabla \phi_i) - \Upsilon_{3,w} \sum_j w_j (\phi_j,\phi_i) + (\mathcal{N}_w, \phi_i). \label{eq:Galsys}
\end{align}
In IGA, the trial functions are chosen from the family of continuous and sufficiently differentiable functions called NURBS~\cite{IGA_book}.

\subsection{B-spline and NURBS}
Let $\displaystyle \Xi = \{\xi_1,\xi_2,\dots,\xi_{m}\}$, with $m = n + p + 1$, be the knot vector, where the knot values $\xi_i \in \mathbb{R}$ are non-decreasing, i.e., $\xi_i \le \xi_{i+1}$, defined in the so-called parameter space, and $p$ be the polynomial order. The knot vector is said to be open if the knot value $\xi_1$ and $\xi_m$ are repeated $p+1$ times. Furthermore, the knot vector is said to be uniform if the knot values partition the parameter space into equal elements.

Using Cox-De Boor's algorithm~\cite{DeBoor}, the B-spline basis functions defined in the parameter space for the knot vector $\Xi$ are given recursively by, for $p = 0$,
\begin{align}
\displaystyle N_{i,0} = \begin{cases}
        &1, ~\text{if}~\xi_i\leq\xi<\xi_{i+1},\\
        &0, ~\text{otherwise},
    \end{cases}
\end{align}
and
\begin{equation}
    N_{i,p} = \frac{\xi - \xi_i}{\xi_{i+p} - \xi_i}N_{i,p-1}(\xi) + \frac{\xi_{i+p+1} - \xi}{\xi_{i+p+1} - \xi_{i+1}}N_{i+1,p-1}(\xi),
\end{equation}
for $p \ge 1$, where $i = 1,2,\dots, n$, with $n$ be the number of B-Spline functions in the basis. Remarks on some properties of the B-Spline functions are in order:
\begin{enumerate}
    \item $N_{i,p}$ are nonnegative, piecewise polynomial functions;
    \item the sum of basis functions for a given order $p$ is identically unity; and
    \item for a uniform knot vector, the functions $N_{i,p} \in C^{p}$. 
\end{enumerate}
The latter property is of importance in the calculation of Greeks, the derivatives of the option price function.

\begin{figure} 
\centering
\includegraphics[width=0.49\textwidth]{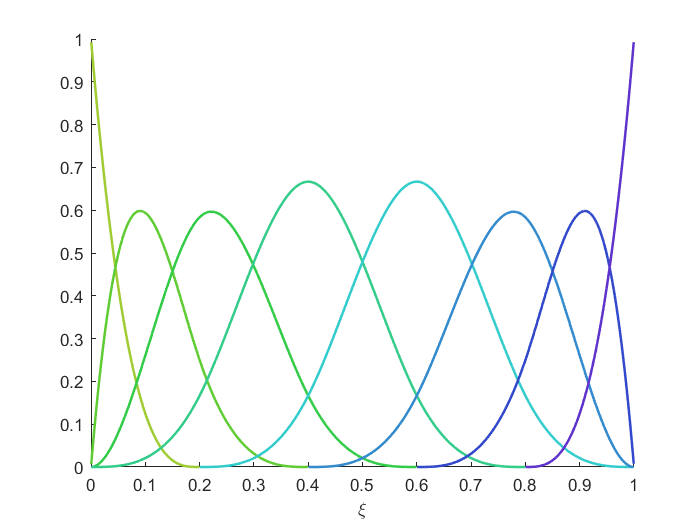}
\includegraphics[width=0.49\textwidth]{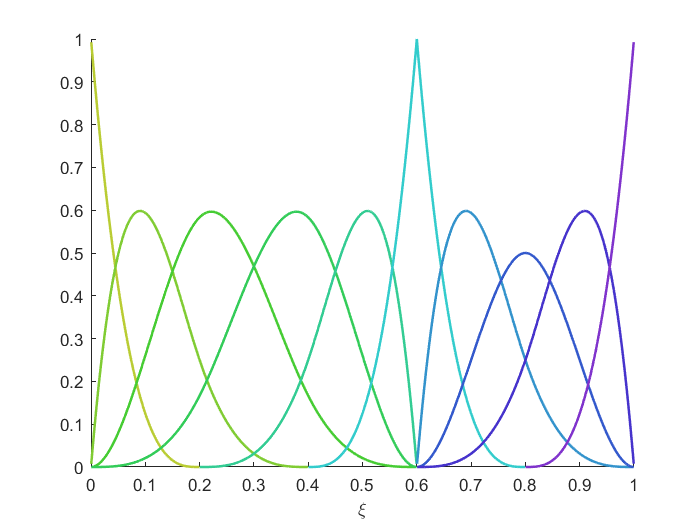}
\caption{Left figure: Cubic B-spline with uniform knot vector as $\Xi = \{0,0,0,0,0.2,0.4,0.6,0.8,1,1,1,1\}$; Right figure: Cubic B-spline with non-uniform knot vector $\Xi = \{0,0,0,0,0.2,0.4,0.6,0.6,0.6,0.8,1,1,1,1\}$.}
\label{fig:Cubic_Bspline}
\end{figure}

Figure~\ref{fig:Cubic_Bspline} shows examples of B-spline basis functions for $p = 3$ (cubic B-spline) on the uniform (left figure) and non-uniform (right figure) open knot vector. In both knot vectors, the first and the last knot are repeated $p+1 = 4$ times, resulting in two basis functions that do not vanish at the end knots. In the uniform case, these are $N_{1,3}$ and $N_{8,3}$, with $N_{1,3}(0) = N_{8,3}(1) = 1$. In the non-uniform case, the basis functions are $N_{1,3}$ and $N_{10,3}$, with $N_{1,3}(0) = N_{10,3}(1) = 1$. Additionally, for the non-uniform knot vector, the knot value $\xi = 0.6$ is repeated 3 times, resulting in basis functions that are non-differentiable at $\xi = 0.6$. Among these non-differentiable basis functions, $N_{6,3}$ attains a value of $1$ at $\xi = 0.6$.



The non-uniform rational B-spline (NURBS) basis functions of order $p$ are defined using the corresponding B-spline basis functions as 
\begin{equation}
    R_{i,p}(\xi) = \frac{\omega_i N_{i,p}(\xi)}{\sum\displaylimits_{i=1}^{n} \omega_i N_{i,p}(\xi)}, \label{eq:NURBSbasis}
\end{equation}
where $\{ \omega_{i} \}_{i=1}^n$ are the NURBS weights. In addition to being piecewise rational functions, NURBS basis functions inherit the properties of B-Spline basis functions listed above. Specifically, setting $\omega_i = \text{constant}$ reduces $R_{i,p}$ to $N_{i,p}$; Thus, B-Spline basis functions are  a special case of NURBS.

\begin{figure} 
\centering
\includegraphics[width=0.49\textwidth]{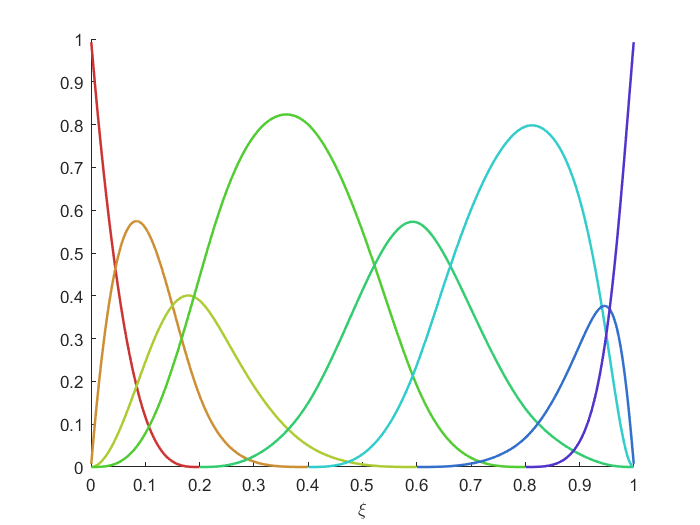}
\includegraphics[width=0.49\textwidth]{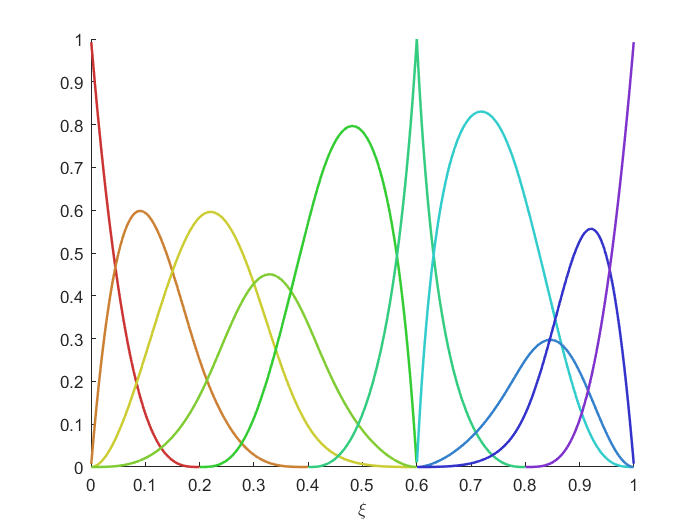}
\caption{Left figure: Cubic NURBS with uniform knot vector and unequal weight vector as $\Xi = \{0,0,0,0,0.2,0.4,0.6,0.8,1,1,1,1\}$, $\omega = \{1,1,1,4,3,5,1,1\}$; Right figure: Cubic NURBS with non-uniform knot vector and unequal weight vector as $\Xi = \{0,0,0,0,0.2,0.4,0.6,0.6,0.6,0.8,1,1,1,1\}$, $\omega = \{1,1,1,1,4,3,     5,1,1,1\}$.}
\label{fig:Cubic_NURBS}
\end{figure}

Examples of NURBS basis functions are shown in Figure~\ref{fig:Cubic_NURBS} for $p = 3$ (cubic NURBS) and the same knot vectors as in B-Splines in Figure~\ref{fig:Cubic_Bspline}. With non-constant weights, the NURBS curves can be viewed as a skewed version of the B-Spline curves.

Consider the finite dimensional subspace $S_0 \subset H^1_0$, where $S_0 = \text{span}\{R_{j,p}\}$, $\forall j \in \{2,3,\dots,n-1\} =  \mathcal{I}_{\Omega}$; Thus, $R_{j,p}$ vanishes at the end knots, with the domain $\Tilde{\Omega} = (0,1)$, defined in the parameter space. The NURBS-based finite-element  approximation to the solution $w$  of~\eqref{eq:unified} using an open knot vector $\Xi$  is the function
$$
  w_h = \sum_{j=2}^{n-1} w_j R_{j,p} + \sum_{j \in \{1,n\}} w_j R_{j,p}.
$$
The Galerkin system~\eqref{eq:Galsys} then reads
\begin{align}
 \frac{\partial}{\partial \tau} \sum_{j=1}^n w_j(R_{j,p},R_{i,p}) &= - \Upsilon_{1,w} \sum_{j=1}^n w_j (\nabla R_{j,p}, \nabla R_{i,p}) - \Upsilon_{2,w} \sum_{j=1}^n w_j (R_{j,p}, \nabla R_{i,p}) - \Upsilon_{3,w} \sum_{j=1}^n w_j (R_{j,p},R_{i,p}) \notag \\
 &+ (\mathcal{N}_w, R_{i,p}), \label{FEMsystem_of_eqxns}
\end{align}
for $i = 2,\dots,n-1$.


\subsubsection{Gauss-Legendre quadrature}
\label{sec:GSQ}
In piecewise polynomial-based FEM, the inner products such as in~\eqref{eq:Galsys} can be evaluated relatively straightforwardly. This is typically done once on a reference element. However, in~\eqref{FEMsystem_of_eqxns}, the complexity of NURBS basis functions and their derivatives makes an analytical evaluation of the inner products impractical. Therefore, numerical integration is employed to approximate these values. A common approach for numerical integration is Gauss-Legendre quadrature~\cite{Legendre1,Legendre2}.

The Gauss-Legendre quadrature for numerical integration of a function $f$ over the interval $[-1,1]$ is
\begin{align}
   \int_{-1}^1 f(\zeta)d\zeta \approx \sum_{i=1}^{p_L} \overline{\omega}_i f(\zeta_i),
\end{align}
where $\zeta_i$ are the zeros of the normalized Legendre polynomial of order $p_L$, $P_{L,p}$, and the integration weights $\overline{\omega}_i$ are given by
$$
\overline{\omega}_i=\frac{b-a}{\left(1-\zeta_i^2\right)\left(P_{L,p}' (\zeta_i)\right)^2}.
$$
To apply the rule on an arbitrary interval $[a,b] \subseteq [\xi_0,\xi_m]$ in, e.g., the parameter space, a change of integration interval to $[-1,1]$ is required, yielding
\begin{align}
  \int_{a}^{b} f(\xi) d\xi = \int_{-1}^1 f\left( \frac{b-a}{2} \zeta + \frac{b+a}{2} \right) \frac{d\xi}{d\zeta} d\zeta,
\end{align}
where $d\xi/d\zeta = (b-a)/2$. For accuracy, the rule is not applied to the entire interval $[\xi_1,\xi_m]$ at once. Instead, the interval is partitioned into subintervals, and the Gauss-Legendre quadrature rule is applied within each subinterval.

\subsubsection{Group FEM for the nonlinear term}
 
To handle the nonlinear term $(\mathcal{N}_w, R_{i,p})$ in ~\eqref{FEMsystem_of_eqxns}, we shall use the group finite-element approach~\cite{FLETCHER1983225}. Assume that the nonlinear function $\mathcal{N}_w$ is approximated by $n$ NURBS basis functions of order $p$:
$$
  \mathcal{N}_w \approx \sum_{j=2}^{n-1} \nu_{j}(w) R_{j,p} + \sum_{j \in \{1,n\}} \nu_{j}(w) R_{j,p}.
$$
Note that the coefficients $\nu_j(w)$ still contain nonlinearity.
The nonlinear term in~\eqref{FEMsystem_of_eqxns} can be written as follows:
\begin{align}
\displaystyle (\mathcal{N}_w,R_{i,p}) =  \sum_{j=2}^{n-1} \nu_j(w) (R_{j,p},R_{i,p}) + \sum_{j \in \{1,n\}} \nu_j(w) (R_{j,p},R_{i,p}),
\end{align}
for all $i = 2,\dots,n-1$. The inner products in the above equation are computed using Gauss-Legendre quadrature rule, outlined in Section~\ref{sec:GSQ}.

\subsubsection{Mapping}
\label{sec: Mapping}
Because the NURBS basis functions are defined in the parameter space, a mapping from the reference interval $\Omega_{\xi} = (\xi_1,\xi_m)$ in the parameter space to the physical space is required. Consider the interval $\Omega = (x_{\min},x_{\max})$ in which the solution of~\eqref{eq:Galsys} is computed. The following mapping of any $x \in \Omega$ can be established:
$$
   x = x_{\min} + \frac{x_{\max} - x_{\min}}{\xi_m - \xi_1}(\xi - \xi_1), \quad \xi \in \Omega_{\xi}.
$$
With $\nabla R_{j,p}(x) = dR_{j,p}(x)/dx = dR_{j,p}(\xi)/d\xi \cdot d \xi/dx$,
\begin{align}
    (\nabla R_{j,p}, \nabla R_{i,p}) = \int_\Omega \frac{d R_{j,p}(x)}{dx} \frac{d R_{i,p}(x)}{dx} dx = \int_{\Omega_{\xi}} \frac{d R_{j,p}(\xi)}{d\xi} \frac{d R_{i,p}(\xi)}{d \xi} \frac{d \xi}{dx} d\xi = \frac{d\xi}{dx} \cdot (\nabla_{\xi} R_{j,p},\nabla_{\xi} R_{i,p})_{\xi},
\end{align}
where $d\xi/dx = |\Omega_{\xi}|/|\Omega|$, and similarly for the other inner products. Substitution of all results to~\eqref{FEMsystem_of_eqxns} gives a system of equations expressed in terms of NURBS basis functions in the reference interval $\Omega_{\xi}$ in the parameter space:
\begin{align}
 \frac{\partial}{\partial \tau} \sum_{j=1}^n w_j \frac{|\Omega|}{|\Omega_{\xi}|}(R_{j,p},R_{i,p})_{\xi} &= - \Upsilon_{1,w} \sum_{j=1}^n w_j \frac{|\Omega_{\xi}|}{|\Omega|}(\nabla_{\xi} R_{j,p}, \nabla_{\xi} R_{i,p})_{\xi} - \Upsilon_{2,w} \sum_{j=1}^n w_j (R_{j,p}, \nabla_{\xi} R_{i,p})_{\xi} \notag \\
 &- \Upsilon_{3,w} \sum_{j=1}^n w_j \frac{|\Omega|}{|\Omega_{\xi}|} (R_{j,p},R_{i,p})_{\xi} \notag + \sum_{j=1}^n \nu_j(w) \frac{|\Omega|}{|\Omega_{\xi}|} (R_{j,p}, R_{i,p})_{\xi}, \label{eq:weaktransform}
\end{align}
for $i = 2,\dots,n-1$.

\section{Time integration scheme}
\label{sec:timeintegral}

Numerical integration of inner products in~\eqref{FEMsystem_of_eqxns} results in the differential algebraic equation
\begin{equation}\label{eq:deagen}
  \frac{\partial}{\partial \tau} (M \mathbf{w} + \mathbf{b}_{M,w}) = - \Upsilon_{1,w} (K \mathbf{w} + \mathbf{b}_{K,w} ) - \Upsilon_{2,w} (N \mathbf{w} + \mathbf{b}_{N,w}) -
  \Upsilon_{3,1} (M \mathbf{w} + \mathbf{b}_{M,w}) 
  + M \mathbf{y}(w) + \mathbf{b}_v =: \mathbf{F}(w),   
\end{equation}
where $\mathbf{w} := [w_j]_{j=2}^{n-1} \in \mathbb{R}^{n-2}$. The matrices $M$, $N$, and $K$ are of size $(n-2)\times (n-2)$ and sparse, due to the small support each NURBS basis function has relative to the computational domain. $\mathbf{b}$ is the boundary condition vector associated with the interpolatory NURBS basis functions $R_{1,p}$ and $R_{n,p}$. The equation~\eqref{eq:deagen} is integrated using the $\theta$-scheme:
\begin{align}
  M\mathbf{w}^{m+1} + \mathbf{b}_{M,w}^{m+1} = M\mathbf{w}^m + \mathbf{b}^m_{M,w} + \Delta \tau \theta \mathbf{F}^{m+1}(w) + \Delta \tau \mathbf{F}^m(w), \label{eq:thetagen}
\end{align}
where $\Delta \tau$ is the time step. To improve stability, a number of Rannacher steps \cite{Rannacher1984} is performed at the initial step from $m =0$ to $m=1$ using the implicit scheme $(\theta = 0)$. 

In the sequel, we discuss the time integration procedure for each model in detail.

\subsection{The Leland model} \label{sec:CN_Leland}

Using the constants given after the Equation~\eqref{eq:unified}, the $\theta$-scheme~\eqref{eq:thetagen} for Leland's model reads
\begin{align}
  A_1 \widehat{\mathbf{v}}^{m+1} - \Delta \tau \theta \left( Le M| \widetilde{\mathbf{v}}^{m+1}| - \mathbf{b}^{m+1}_{M,\widetilde{v}}\right) &= A_0 \widehat{\mathbf{v}}^{m}  - \Delta \tau (1-\theta) \left( Le M| \widetilde{\mathbf{v}}^{m}| - \mathbf{b}^{m}_{M,\widetilde{v}}\right) \notag \\
  &- \Delta \tau \theta \left( \mathbf{b}^{m+1}_{K,\widehat{v}} + \mathbf{b}^{m+1}_{N,\widehat{v}} \right) - \Delta \tau (1-\theta) \left( \mathbf{b}^{m}_{K,\widehat{v}} + \mathbf{b}^m_{N,\widehat{v}} \right), \label{eq:thetaleland}
\end{align}
where $A_0 = M - \Delta \tau (1-\theta)(K + N)$ and $A_1 = M + \Delta \tau \theta (K + N)$, and $\widehat{\mathbf{v}}^m := [\widehat{v}_j]$ satisfies the equation
$$
  M \widetilde{\mathbf{v}}^m = - (K + N) \widehat{\mathbf{v}}^m - \mathbf{b}^m_{K,\widehat{v}} - \mathbf{b}^m_{N,\widehat{v}},
$$
obtained from the weak formulation of the second PDE in~\eqref{eq:system_of_transformed_eqxns_for_Leland}. Linearization of~\eqref{eq:thetaleland} by assuming $|\widetilde{\mathbf{v}}^{m+1}| \approx |\widetilde{\mathbf{v}}^{m}|$ results in the time integration procedure
\begin{align}
\begin{cases}
  \widetilde{\mathbf{v}}^m = - M^{-1} \left((K + N) \widehat{\mathbf{v}}^m + \mathbf{b}^m_{K,\widehat{v}} - \mathbf{b}^m_{N,\widehat{v}} \right), \\
  \mathbf{f}^m = A_0 \widehat{\mathbf{v}}^m - \Delta \tau \left( Le M| \widetilde{\mathbf{v}}^{m}| - \mathbf{b}^{m}_{M,\widetilde{v}}\right) - \Delta \tau \theta \left( \mathbf{b}^{m+1}_{K,\widehat{v}} + \mathbf{b}^{m+1}_{N,\widehat{v}} \right) - \Delta \tau (1-\theta) \left( \mathbf{b}^{m}_{K,\widehat{v}} + \mathbf{b}^m_{N,\widehat{v}} \right), \\
  \widehat{\mathbf{v}}^{m+1} = A_1^{-1} \mathbf{f}^m,
\end{cases}
\label{eq:DAE Leland model}
\end{align}
for $m =0,1,\dots.$

\subsection{The AFV model} \label{sec:CN_AFV}


The $\theta$-scheme for the AFV model reads
\begin{align}
    A_{11} \bfU^{m+1} & - \rho \theta \Delta \tau M\left(P_{put}^{m+1} (\bfU_{put}^{\star,m+1} - \bfU^{m+1}) + P_{call}^{m+1}(\bfU^{m+1}-\bfU^{\star,m+1}_{call})\right) 
    \notag \\[3pt]
    &= \widetilde{A}_{11}\bfU^m - \theta \Delta \tau p M \delta^{m+1} -(1-\theta) \Delta \tau p M \delta^m \notag \\[3pt]
    &+ \rho (1-\theta) \Delta \tau M\left(P_{put}^{m} (\bfU_{put}^{\star,m} - \bfU^{m}) + P_{call}^{m}(\bfU^{m}-\bfU^{\star,m}_{call})\right) + \theta \Delta \tau \boldsymbol{\beta}_1^{m+1} +  (1-\theta) \Delta \tau \boldsymbol{\beta}_1^{m} \notag \\[3pt]
    &+ \hat{\bfb}^m_{M,U} - \hat{\bfb}^{m+1}_{M,U} + \theta \rho \Delta \tau (\bfb_{put}^{m+1} + \bfb_{call}^{m+1}) + (1-\theta) \rho \Delta \tau (\bfb_{put}^{m} + \bfb_{call}^{m}), \label{CNsystem1} \\[3pt]
    A_{22} \bfB^{m+1} &= \widetilde{A}_{22}\bfB^m + \theta \Delta \tau \boldsymbol{\beta}_2^{m+1} +  (1-\theta) \Delta \tau \boldsymbol{\beta}_2^{m} + \hat{\bfb}^m_{M,B} - \hat{\bfb}^{m+1}_{M,B}, \label{CNsystem2}\\[3pt]
    A_{33} \bfC^{m+1} &= \widetilde{A}_{33}\bfC^m - \theta \Delta \tau p M \gamma^{m+1} -(1-\theta) \Delta \tau p M \gamma^m+ \theta \Delta \tau \boldsymbol{\beta}_3^{m+1} +  (1-\theta) \Delta \tau \boldsymbol{\beta}_3^{m} + \hat{\bfb}^m_{M,C} - \hat{\bfb}^{m+1}_{M,C}, \label{CNsystem3}
\end{align}
where $P_{put} = \text{diag}(\alpha_{put,j})$ and $P_{call} = \text{diag}(\alpha_{call,j})$, with $\alpha_{put,call}$ given in~\eqref{eq:alpha}. Furthermore, in~\eqref{CNsystem1}--\eqref{CNsystem3},
\begin{align}
    A_{11} &= M + \theta \Delta \tau \left( \frac{\sigma^2}{2} K + \left(r+p\eta - \frac{\sigma^2}{2}  \right)N + (r+p) M  \right), \notag \\
    A_{22} &= M + \theta \Delta \tau \left( \frac{\sigma^2}{2} K + \left(r+p\eta - \frac{\sigma^2}{2}  \right)N + (r+p) M +\hat{R}pM \right), \notag \\
     A_{33} &= M + \theta \Delta \tau \left( \frac{\sigma^2}{2} K + \left(r+p\eta - \frac{\sigma^2}{2}  \right)N + (r+p) M\right), \notag \\
    \widetilde{A}_{11} &= M - (1-\theta) \Delta \tau\left( \frac{\sigma^2}{2} K + \left(r+p\eta - \frac{\sigma^2}{2}  \right)N + (r+p) M  \right), \notag \\
    \widetilde{A}_{22} &= M - (1-\theta) \Delta \tau \left( \frac{\sigma^2}{2} K + \left(r+p\eta - \frac{\sigma^2}{2}  \right)N + (r+p) M +\hat{R}pM \right), \notag\\
     \widetilde{A}_{33} &= M - (1- \theta) \Delta \tau \left( \frac{\sigma^2}{2} K + \left(r+p\eta - \frac{\sigma^2}{2}  \right)N + (r+p) M\right), \notag \\
    \displaystyle \boldsymbol{\beta}_1 &= \frac{\sigma^2}{2}  \bfb_{K,U} + \left(r +p\eta- \frac{\sigma^2}{2}  \right)\bfb_{N,U} +(r+p) \bfb_{M,U} - p\bfb_{M,\delta}, \notag \\
    \displaystyle \boldsymbol{\beta}_2   &= \frac{\sigma^2}{2} \bfb_{K,B} + \left(r+p\eta - \frac{\sigma^2}{2} \right)\bfb_{N,B} + (r+p)\bfb_{M,B} - \hat{R}p\bfb_{M,B}, \notag \\
    \displaystyle \boldsymbol{\beta}_3  &= \frac{\sigma^2}{2} \bfb_{K,C} + \left(r+p\eta - \frac{\sigma^2}{2} \right)\bfb_{N,C} + (r+p)\bfb_{M,C} - p\bfb_{M,\gamma}. \notag
\end{align}
Recall that $\delta$ and $\gamma$ are functions of $\bfB$.

Suppose the solutions $\bfU^m$, $\bfB^m$, and $\bfC^m$ are known. The solutions at the next time level $m+1$ are  computed by solving~\eqref{CNsystem2} for $\bfB^{m+1}$, without applying the constraints. With $\bfB^{m+1}$, $\bfC^{m+1}$ is computed from~\eqref{CNsystem3}. The constraints are then applied to both $\bfB^{m+1}$ and $\bfC^{m+1}$. The solution $\bfU^{m+1}$ is then computed via~\eqref{CNsystem1} using the known $\bfU^{m}$, $\bfB^{m}$, and the constrained $\bfB^{m+1}$.

\subsubsection{Boundary solutions}

The solution procedure outlined earlier requires, however, solutions at the boundaries at the time level $m+1$. 
At $x_{\min}$, with $U_0(\tau) := U(x_{\min},\tau)$, $B_0(\tau) := B(x_{\min},\tau)$, and $C_0(\tau) := C(x_{\min},\tau)$, the boundary conditions can be written as follows:
\begin{eqnarray}
\begin{cases}
\displaystyle \frac{\partial U_0(\tau)}{\partial \tau} =  -(r+p)U_0(\tau) - p\hat{R}B_0(\tau),\\[10pt]
\displaystyle \frac{\partial B_0(\tau)}{\partial t} = -(r+(1-\hat{R})p) B_0(\tau),\\[10pt]
\displaystyle \frac{\partial C_0(\tau)}{\partial t} = -(r+p) C_0(\tau).
\end{cases} \label{bcond0a}
\end{eqnarray}
Application of the $\theta$-scheme on \eqref{bcond0a} leads to the discrete equations:
\begin{align}
    U_0^{m+1} + \theta \Delta \tau \left( (r+p) U_0^{m+1} + \hat{R}p V_0^{m+1}\right) &= U_0^m - (1-\theta) \Delta \tau \left((r+p) U_0^{m} + \hat{R}p V_0^{m} \right), \label{CNbcond1} \\
    \left(1 + \theta \Delta \tau (r+(1-\hat{R})p)\right) V_0^{m+1} &= \left(1 - (1-\theta) \Delta \tau (r+(1-\hat{R})p)\right) V_0^m, \label{CNbcond2} \\
    \left(1 + \theta \Delta \tau (r+p)\right) C_0^{m+1} &= \left(1 - (1-\theta) \Delta \tau (r+p)\right) C_0^m. \label{CNbcond3}
\end{align}
For the boundary solutions, $V_0^{m+1}$ and $C_0^{m+1}$ are computed independently. The solution $V_0^{m+1}$ is then used to compute $U_0^{m+1}$.

\subsubsection{Interior solutions}
With solutions at the boundaries available at $\tau^m$ and $\tau^{m+1}$, all related boundary vectors in~\eqref{CNsystem1}--~\eqref{CNsystem3} are known. We are thus now in the position to compute the solutions $\bfU^{m+1}$, $\bfB^{m+1}$, and $\bfC^{m+1}$. First, $\bfB^{m+1}$ is computed from~\eqref{CNsystem2}. With $\bfB^{m+1}$ known, $\bfC^{m+1}$ can be computed from~\eqref{CNsystem3}. Then we apply the inequality constraints on $\bfB^{m+1}$ and $\bfC^{m+1}$. With the already adjusted $\bfB^{m+1}$ known, the equation~\eqref{CNsystem1} reduces to a nonlinear function of $\bfU^{m+1}$. To deal with the nonlinearity, we approximate the penalty terms in~\eqref{CNsystem1} in a fully implicit way, resulting in the equation
\begin{eqnarray}
   0 &= A_{11} \bfU^{m+1} - \rho \Delta \tau M\left(P_{put}^{m+1} (\bfU_{put}^{\star,m+1} - \bfU^{m+1}) + P_{call}^{m+1}(\bfU^{m+1}-\bfU^{\star,m+1}_{call})\right) 
    - \boldsymbol{\phi} := \bsf(\bfU^{m+1}), \label{eqnonlinU}
\end{eqnarray}
where
\begin{align}
   \boldsymbol{\phi} &= \widetilde{A}_{11}\bfU^m + \widetilde{A}_{12} \bfB^m  - A_{12} \bfB^{m+1} + \theta \Delta \tau \boldsymbol{\beta}_1^{m+1} +  (1-\theta) \Delta \tau \boldsymbol{\beta}_1^{m} + \widehat{\bfb}^m_{M,U} - \widehat{\bfb}^{m+1}_{M,U} + \theta \Delta \tau (\bfb_{put}^{m+1} + \bfb_{call}^{m+1})\notag \\
    &- \theta \Delta \tau p M \delta^{m+1} -(1-\theta) \Delta \tau p M \delta^m. \notag
\end{align}

The nonlinear equation~\eqref{eqnonlinU} is solved iteratively using Newton's method. Starting from an initial guess of the solution $\bfU^{m+1,0}$, the solution $\bfU^{m+1}$ is approximated using the iterands
$$
   \bfU^{m+1,k} = \bfU^{m+1,k-1} - \left( \nabla \bsf (\bfU^{m+1,k-1})\right)^{-1} \bsf(\bfU^{m+1,k-1}), \quad k = 1, 2, \dots,
$$
where $\nabla \bsf(\bfU^{m+1,k-1})$ is the Jacobian of $\bsf$, given by
\begin{eqnarray}
  \nabla \bsf(\bfU^{m+1,k-1}) = A_{11} + \rho \Delta \tau M \left(P_{call}^{m+1,k-1} - P_{put}^{m+1,k-1} \right).  \label{eqjacf}
\end{eqnarray}
The initial guess $\bfU^{m+1,0}$ is chosen such that it solves unconstrained AFV PDE, which is equivalent to solving~\eqref{CNsystem1} without penalty terms. We shall also use this unconstrained solution $\bfU^{m+1,0}$ to constrain the initially computed $\bfB^{m+1}$ prior to the start of Newton's iterations. 


The procedure for computing the solutions in the interior for one $\theta$-scheme step is summarized in Algorithm~\ref{alg:Interior}.

\begin{algorithm}[H] 
\begin{algorithmic}[1] 
\Require input $\bfU^m$, $\bfB^m$;
\State compute $B_p(\tau^{m+1})$ and $B_c(\tau^{m+1})$;
\State compute $\bfB^{m+1}$ from~\eqref{CNsystem2}, without constraints;
\State compute $\gamma$;
\State compute $\bfC^{m+1}$ from~\eqref{CNsystem3};
\State apply minimum-maximum value constraints (i.e., \eqref{constraint:v_constrain_1} and~\eqref{constraint:v_constrain_2}) on $\bfB^{m+1}$ using $\bfC^{m+1}$;
\For{$k = 1,2,\dots$ until convergence}
    \State compute $\bsf(\bfU^{m+1,k-1})$ using~\eqref{eqnonlinU};
    \State compute $\nabla \bsf(\bfU^{m+1,k-1})$ using~\eqref{eqjacf};
    \State $\bfU^{m+1,k} \leftarrow \bfU^{m+1,k-1} - \left( \nabla \bsf(\bfU^{m+1,k-1}) \right)^{-1} \bsf(\bfU^{m+1,k-1})$;
    \If {$\parallel\bfU^{m+1,k} - \bfU^{m+1,k-1}\parallel_{\infty}\leq tol~\text{OR}~\left[P_{call}^{m+1,k-1} = P_{call}^{m+1,k}~\text{AND}~ P_{put}^{m+1,k-1} = P_{put}^{m+1,k}\right]$} 
        \State\algorithmicbreak
    \EndIf
\EndFor
\State apply the constraints ~\eqref{constraint:c_constrain_1} and~\eqref{constraint:c_constrain_2} on  $\bfB^{m+1}$ using $\bfU^{m+1,k}$;
\If{$\tau^{m+1} \in \mathcal{T}_{\text{coupon}}$}
    \State $\bfU^{m+1} \leftarrow \bfU^{m+1} + K_{coup}$;
    \State $\bfB^{m+1} \leftarrow \bfB^{m+1} + K_{coup}$;
\EndIf
\end{algorithmic}
\caption{Computing the interior solutions}\label{alg:Interior}
\end{algorithm}

\section{Numerical results}
\label{sec:numresults}
In this section, we present numerical solutions of the Leland and AFV model using IGA, described in Sections~\ref{sec:methodology} and~\ref{sec:timeintegral}. In all cases, we use cubic NURBS IGA, as it offers some advantages over the other choice of NURBS basis functions in the literature~\cite{IGA_book,Hughes2005}. Because there is no exact solution, to gauge the accuracy of our numerical solutions, we compare the IGA solutions with FDM, P1-FEM, and P2-FEM~\cite{Wei2024,Ankudinova2008,Ayache2003}. 


\subsection{Linear Black-Scholes model}\label{sec: Linear Black Scholes}

\begin{figure}[H]
\centering
\includegraphics[width=0.49\textwidth]{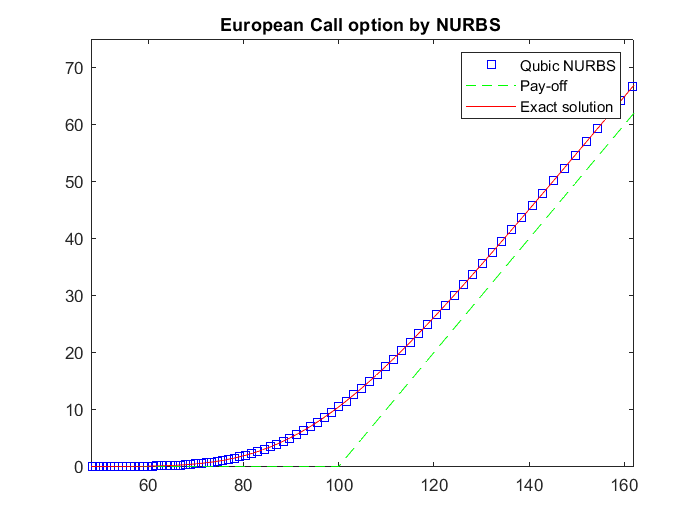}
\includegraphics[width=0.49\textwidth]{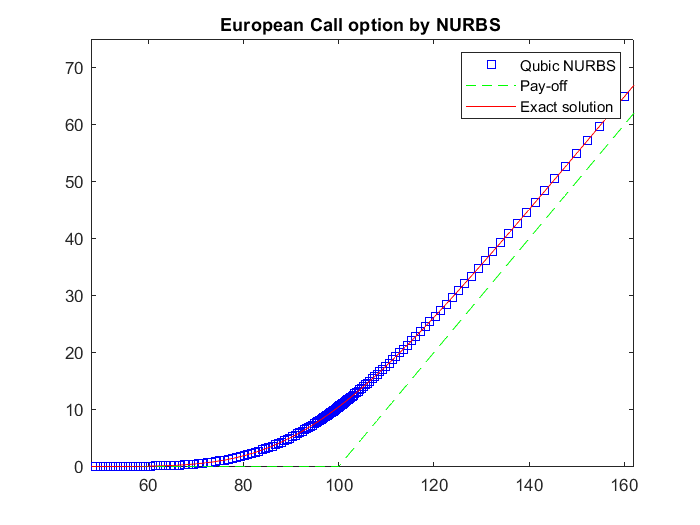}
\caption{Solution of the European call option at $t = 0$, computed using cubic NURBS-IGA with constant weights (thus, cubic B-Splines) with $r = 0.05$, $\sigma = 0.2$, $\hat{K} = \$100$. Left figure: Uniform knots with $nE = 2^8$, $n_{\tau} = 6\times 10^4$; Right figure: Non-uniform knots with $nE = 2^8$, $n_{\tau} = 6\times 10^4$.}
\label{fig:Linear_BS_nonuniform_larger_scale}
\end{figure}

For benchmarking, we first consider the European call option pricing modeled by the linear Black-Scholes equation. In this case, an exact solution exists. We numerically solve the linear PDE model using cubic NURBS-based IGA with constant weights (thus, cubic B-Splines) and with both uniform and non-uniform knot vectors. The non-uniform knot vector approach is employed to more accurately approximate the initial condition, which is a piecewise continuous function with a kink at the strike price $\hat{K} = 100$. For the non-uniform knot vectors, denser knots are placed around the kink with $\xi = 1/2$, $3 \times \text{repeated}$ knots exactly at the kink. The IGA solutions at the initial time $t = 0$ are shown in Figure~\ref{fig:Linear_BS_nonuniform_larger_scale}, for the uniform and the non-uniform knot vectors corresponding to $2^8$ elements. In both cases, the numerical solution agrees well with the exact solution. For instance, at $S = 100$,  the solution using the uniform and nonuniform knot are $U(100) = 10.4835$ and $U(100) = 10.4513$, respectively, the latter is very close to the exact solution $\$10.4505$ than the former.

\begin{figure}[h!]
\centering
\includegraphics[width=0.49\textwidth]{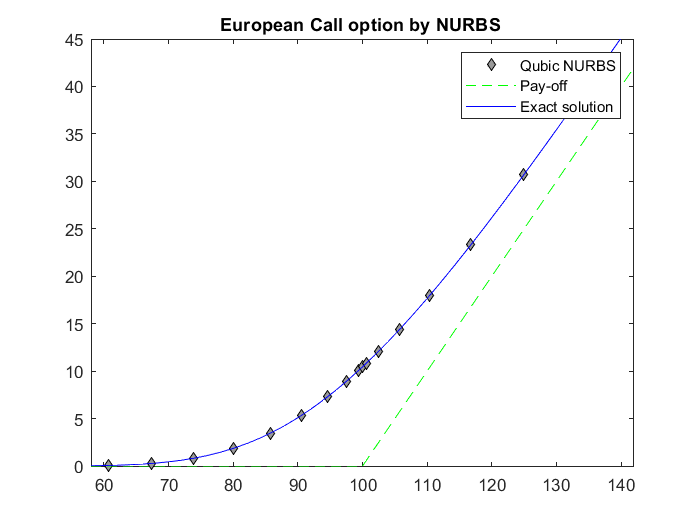}
\includegraphics[width=0.49\textwidth]{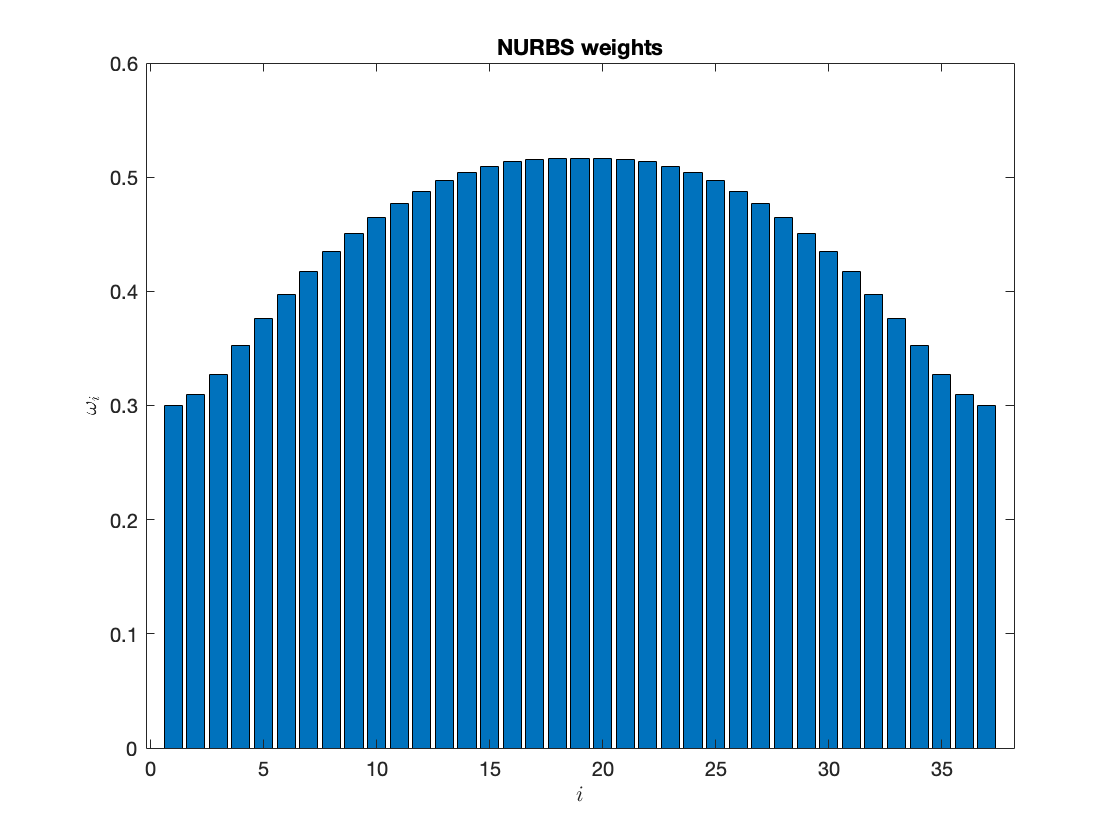}
\caption{Solution of the European call option at $t = 0$, with $r = 0.1$, $\sigma = 0.2$, and $\hat{K} = \$100$, computed using nonuniform, fitted NURBS with $nE = 2^5$ and $n_\tau = 6 \times 10^4$ (left figure). The weights are given in the right figure.}
\label{fig:Linear_BS_nonuniform_fitted}
\end{figure}

Figure~\ref{fig:Linear_BS_nonuniform_fitted} (left) compares the cubic NURBS-IGA solution with  nonuniform weights with the the exact solution. The IGA solution is obtained using $2^5$ nonuniform knots and weights, which are shown in Figure~\ref{fig:Linear_BS_nonuniform_fitted} (right). As observed from the Figure, the solution of the weighted cubic NURBS-IGA agrees well with the exact solution.

The superiority of weighted cubic NURBS-IGA over unweighted cubic NURBS (B-Splines) is shown in Table~\ref{tab: LinearbS_nonuniform_new}, where the convergence of the computed solution at $S  100$ towards the exact solution is demonstrated. Clearly not only that the weighted NURBS-IGA converges extremely fast, but this fast convergence is achieved by using a much smaller number of knots, which in turn reduces the computational time. Nonuniform B-Splines-based IGA tends to converge faster than the uniform IGA; Interestingly, its convergence is no faster than P2-FEM. We note here, however, that for P2-FEM, the number of degree of freedom (i.e., the number of basis functions needed to construct the approximate solution) is $2n_E - 1$. In comparison, for  uniform knot vectors, the number of cubic NURBS basis functions to approximate the solution in the interior is $n_E+3$; a nonuniform knot vector ends up in approximately $n_E$ basis functions. Thus, for the same number of elements, cubic NURBS requires nearly half the computational work per time step of P2-FEM.

\begin{table}[h!]
\caption{European call option price at $t=0$ and $\hat{K} = \$100$, denoted by $U(100)$, computed by NURBS. The exact price is $\$10.4505$. The error is the absolute difference between the exact and numerical solution. In all computations, $n_\tau = 6 \times 10^4$.} \label{tab: LinearbS_nonuniform_new}
\begin{center}
\begin{tabular}{c|cc|cccc|cc}\hline
& & & \multicolumn{4}{c|}{Unweighted cubic NURBS} & \multicolumn{2}{c}{Weighted cubic NURBS} \\ \cline{4-9} 
  $nE$ & \multicolumn{2}{c|}{P2-FEM} &  \multicolumn{2}{c}{Uniform} & \multicolumn{2}{c|}{Nonuniform} & \multicolumn{2}{c}{Nonuniform}  \\ 
       & $U(100)$ & Error & $U(100)$ & Error & $U(100)$ & Error & $U(100)$ & Error \\ \hline \hline
  $2^5$ & 10.3792 & 0.0712 &12.2987 & 1.8482 &10.5256 &0.0751 & 10.4505 & 0.0000 \\
      $2^6$ & 10.4702& 0.0197 & 10.9524& 0.5019 & 10.4691& 0.0185& 10.4505 & 0.0000 \\
 $2^7$ & 10.4513 & 0.0008 &10.5652 & 0.1147 & 10.4544 &   0.0025&  10.4505 & 0.0000 \\
 $2^{8}$ & 10.4506 & 0.0001 &10.4835 & 0.0330 & 10.4513 &  0.0006&  10.4505 & 0.0000 \\
$2^{9}$ &  10.4505 & 0.0000 &10.4555 & 0.0050 &10.4507 &  0.0002 &   &  \\
 $2^{10}$ &  10.4505 & 0.0000 &10.4519 & 0.0014 &10.4505 &  0.0000 &   &  \\
 $2^{11}$ &  10.4505 & 0.0000 &10.4509 & 0.0004 & 10.4505 &  0.0000 &  &  \\
 $2^{12}$ &  10.4505 & 0.0000 &10.4507 & 0.0002 & 10.4505 &  0.0000 &  &  \\
\hline
\end{tabular}
\end{center}
\end{table}

As for the linear Black-Scholes PDE, the use of weighted cubic NURBS results in fast convergence to the solution $U(100) = 10.4505$ using only $2^5$ elements; see Table \ref{tab: LinearbS_nonuniform_new}. To obtain this result, we use the weights as shown in Figure~\ref{fig:Linear_BS_nonuniform_fitted}. Increasing the number of elements does not change the solution within the first 4 fractional digits. As we refine the mesh, we however, have to adjust the weights in order to achieve this mesh-independent convergence. While it retains a similar distribution as for $nE = 2^5$, the maximum weight in the two cases is different and to be adjusted accordingly.

\subsection{The Leland model}

We consider numerical solutions of the transformed Leland PDE with parameters $r = 0.1$, $\sigma = 0.2$, $\hat{K} = \$100$, $T = 1$, and $Le \approx 0.8$ and $ 1.3$. The choice $Le > 1$ results in an ill-posed problem and hence poses challenges when computing the solution numerically. For numerical stability, we shall set the time step $\Delta \tau$ and the spatial mesh $\Delta x$ such that $\Delta \tau/\Delta x < 1$ and $\Delta \tau/\Delta x^2 < 1$ as in~\cite{Wei2024}. 

\begin{figure}[H]
\centering
\includegraphics[width=0.46\textwidth]{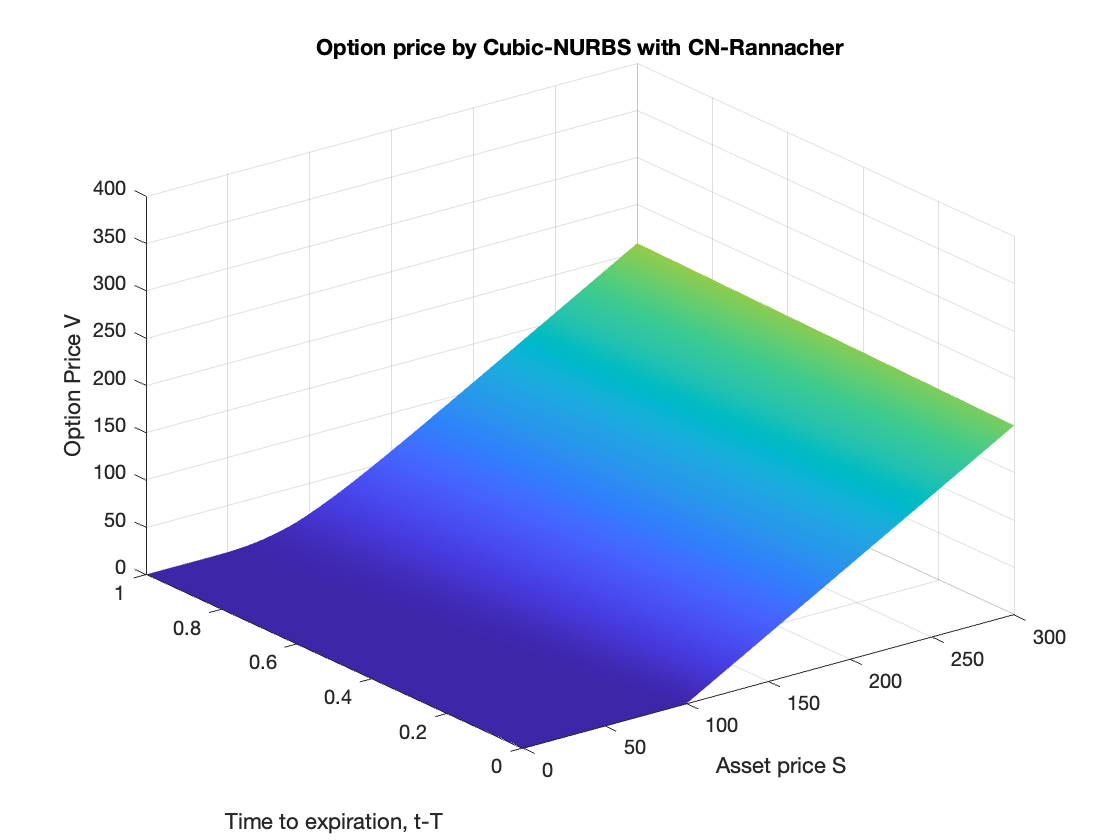}
\includegraphics[width=0.46\textwidth]{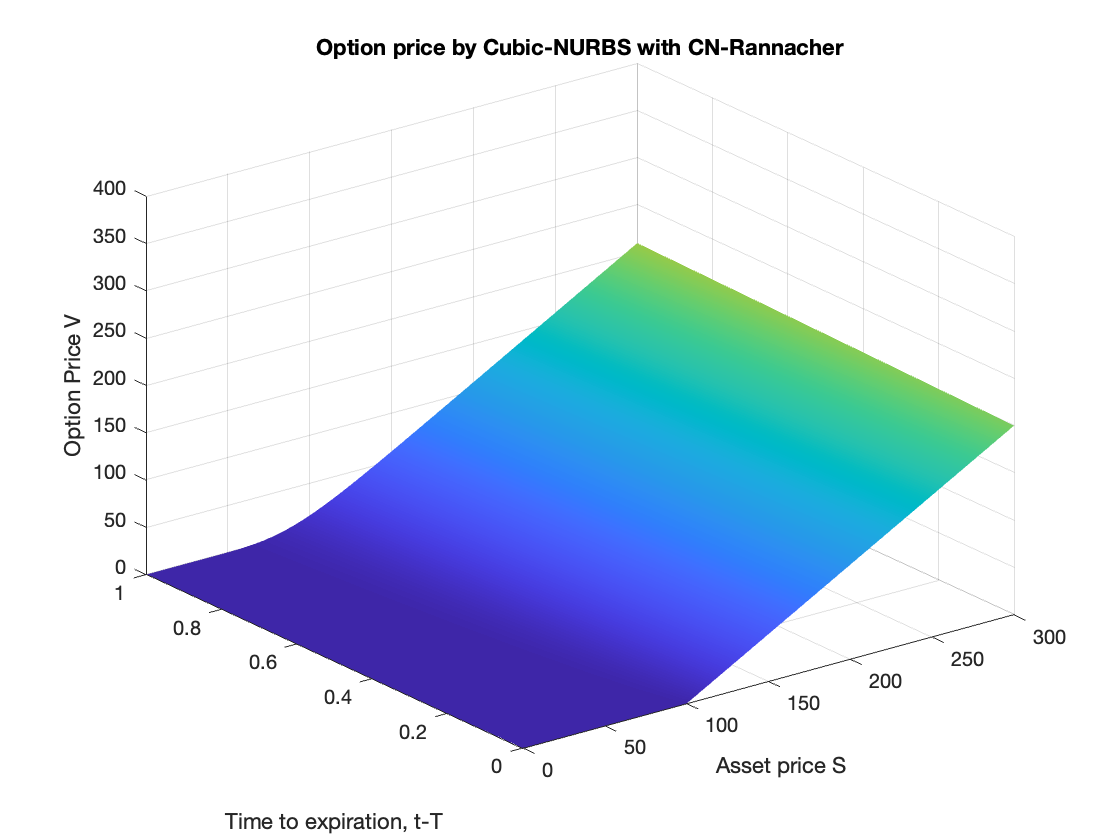}
\caption{Solution of the Leland model calculated using unweighted cubic NURBS IGA on uniform knots, with $r = 0.1$, $\sigma = 0.2$, $\hat{K} = \$100$, and $Le \approx 0.8$. Left: $\Delta\tau/\Delta x = 0.02$, $\Delta\tau/\Delta x^2 = 0.8$; Right: $\Delta\tau/\Delta x = 0.05$, $\Delta\tau/\Delta x^2 = 0.1$.}
\label{fig:Leland_surface}
\end{figure}
\begin{figure}[H]
\centering
\includegraphics[width=0.47\textwidth]{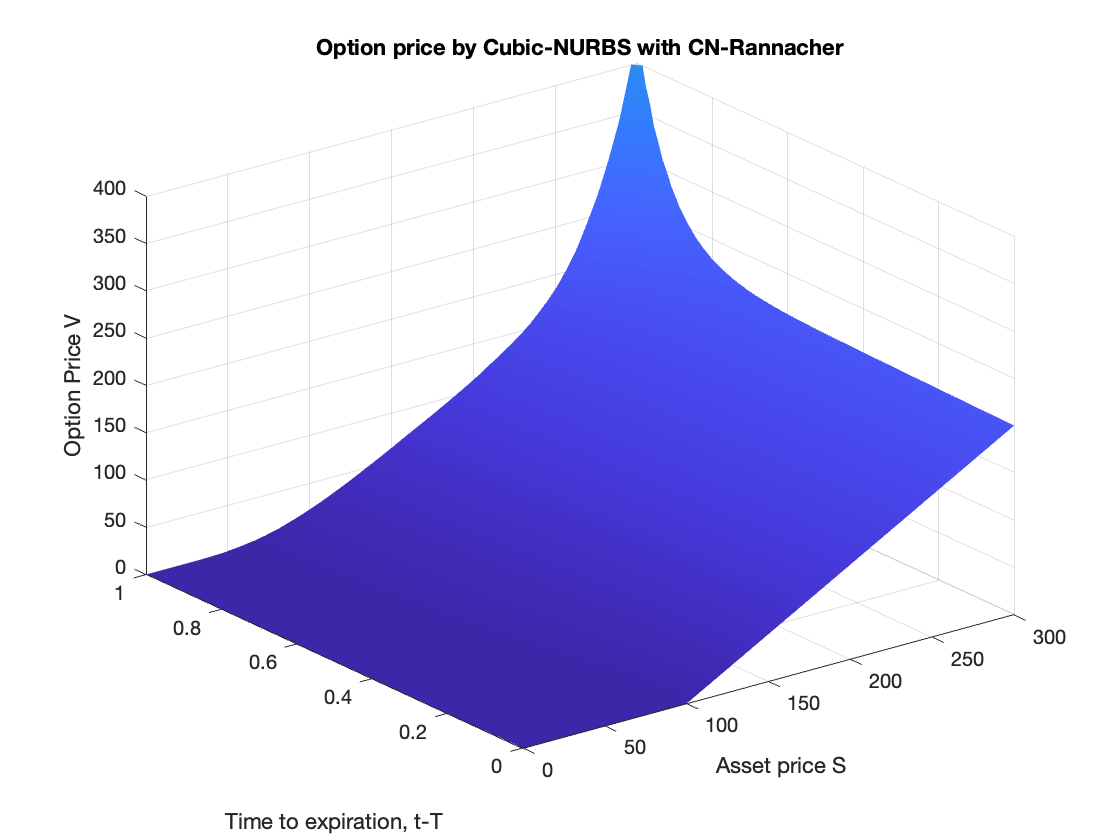}
\includegraphics[width=0.47\textwidth]{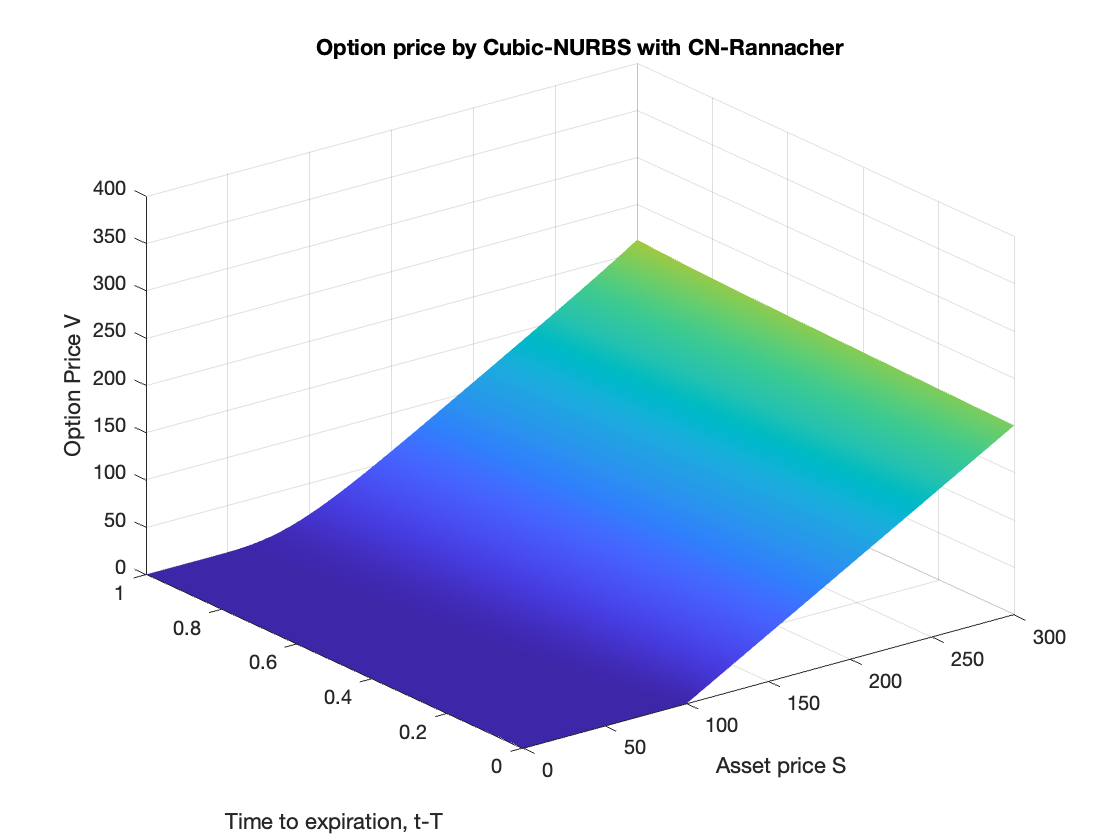}
\caption{Solution of the Leland model calculated using unweighted cubic NURBS IGA on uniform knots, with $r = 0.1$, $\sigma = 0.2$, $\hat{K} = \$100$, and $Le \approx 1.33$. Left: $\Delta\tau/\Delta x = 0.02$, $\Delta\tau/\Delta x^2 = 0.8$; Right: $\Delta\tau/\Delta x = 0.05$, $\Delta\tau/\Delta x^2 = 0.1$.}
\label{fig:Leland_surface_133_values}
\end{figure}

Numerical solutions for $Le \approx 0.8$ in the $t$-$S$ space are shown in Figure~\ref{fig:Leland_surface}, calculated using two discretization settings (details provided in the figure). The two solution surfaces suggest that the cubic NURBS-based IGA is a stable method for this problem. The solution surfaces for $Le \approx 1.3$ (an ill-posed problem) are shown in Figure~\ref{fig:Leland_surface_133_values}. When a coarse time-space mesh is used, the solution exhibits instability near the boundary as $t$ approaches 0, resulting in a blow-up. Notably, this near-boundary instability is absent in the P1-FEM and FDM solutions (not shown here). For the cubic NURBS IGA, this instability can, however, be mitigated by by adopting finer temporal-spatial discretization, as demonstrated in (Figure~\ref{fig:Leland_surface_133_values}: Right).

\begin{figure}[H]
\centering
\includegraphics[width=0.49\textwidth]{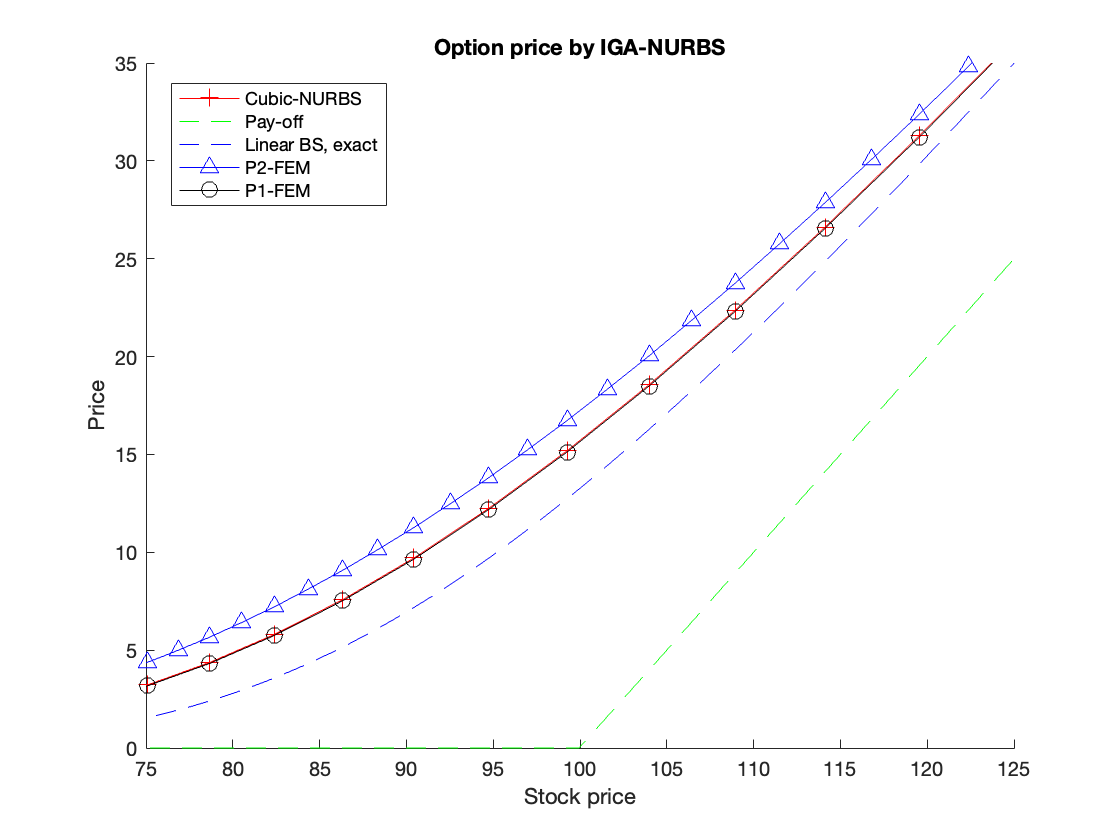}
\includegraphics[width=0.49\textwidth]{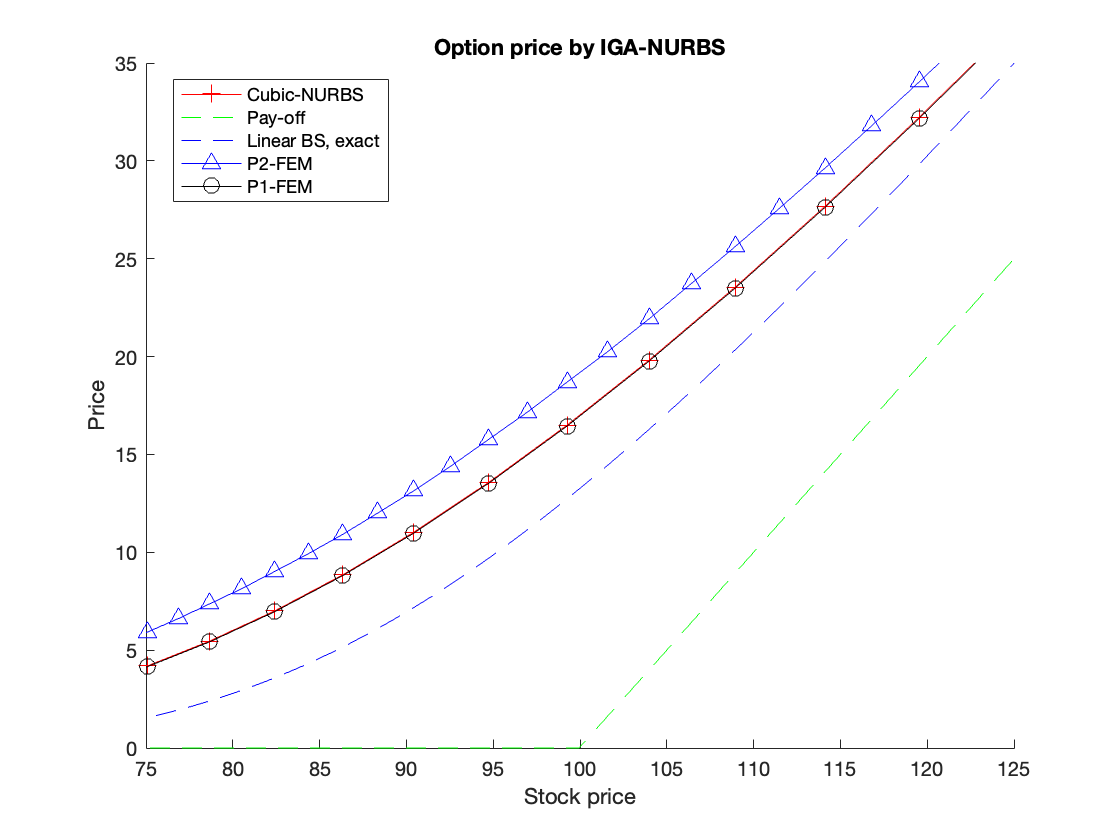}
\caption{Solution of the Leland model $t = 0$, $r = 0.1$, $\sigma = 0.2$, $\hat{K} = \$100$. Left figure: $Le \approx 0.8$, $\Delta x = 0.04$, $\Delta\tau = 0.00025$, therefore, $\Delta\tau/\Delta x = 0.005$, $\Delta\tau/\Delta x^2 = 0.1$; Right figure: $Le \approx 1.33$, $\Delta x = 0.04$, $\Delta\tau = 0.00025$, therefore, $\Delta\tau/\Delta x = 0.005$, $\Delta\tau/\Delta x^2 = 0.1$.}
\label{fig:2D_plots_Leland_by_NURBS}
\end{figure}

In Figure~\ref{fig:2D_plots_Leland_by_NURBS}, we compare the cubic NURBS solutions with P1- and P2-FEM, for $Le \approx 0.8$ (left figure) and $Le \approx 1.3$ (right figure) at $t = 0$, computed using very fine time steps and elements. We note that the P1-FEM solution agrees well with the FDM solution. In the absence of an exact solution, the P1-FEM solution will be used as a reference solution. As observed in the figure, the plot of the cubic NURBS solution is practically indistinguishable from the P1-FEM solution, indicating a good match between the two solutions.

\begin{figure}[H]
\centering
\includegraphics[width=0.49\textwidth]{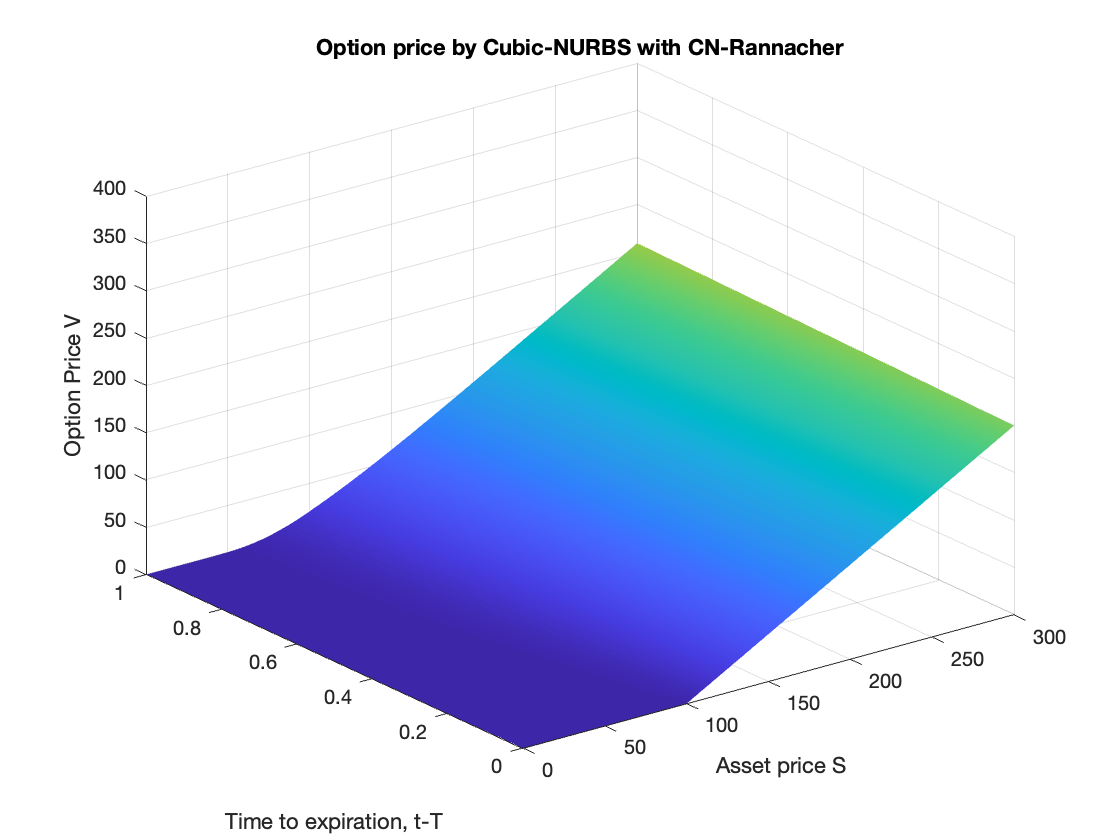}
\includegraphics[width=0.49\textwidth]{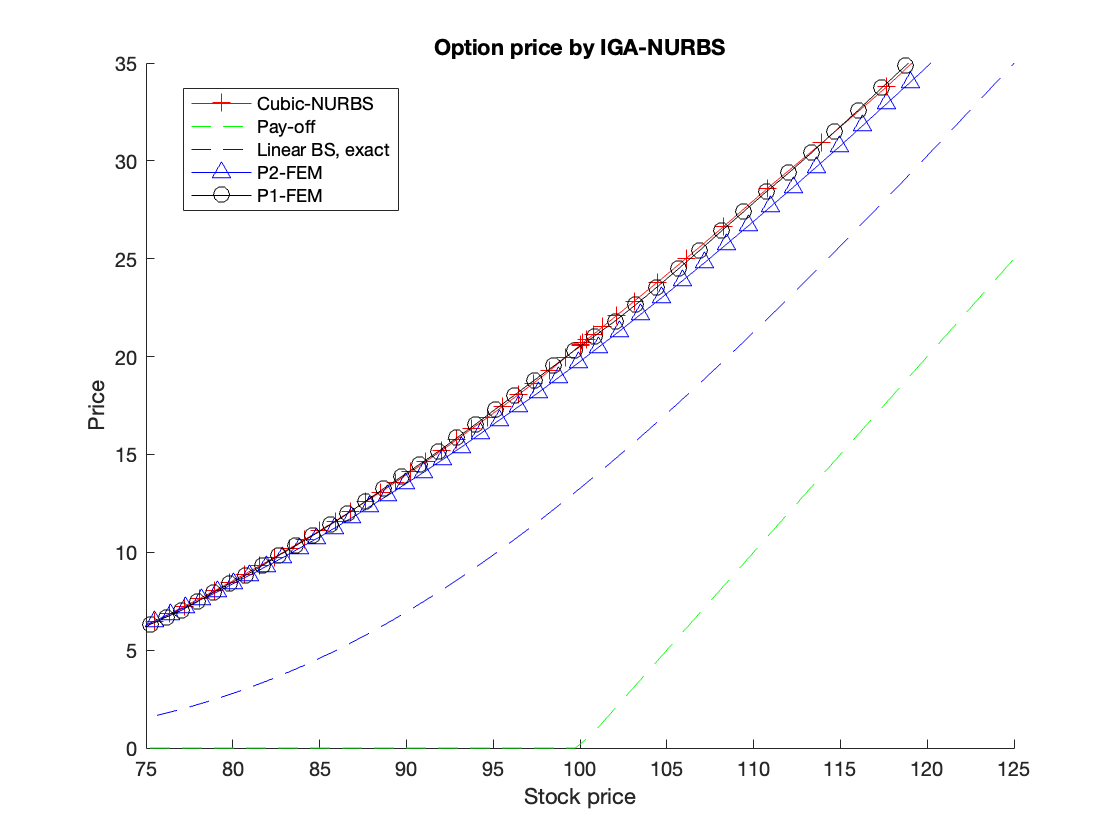}
\caption{Solution of the Leland model calculated using weighted cubic NURBS IGA on $2^8$ nonuniform knots, with $r = 0.1$, $\sigma = 0.2$, $\hat{K} = \$100$, and $Le \approx 0.8$. For P1-FEM and P2-FEM, $2^9$ and $2^8$ elements are used, respectively.}
\label{fig:Leland_surface_slice_using_Weighted_nonuniform_NURBS}
\end{figure}

Next, we solve the Leland model using weighted cubic NURBS on $2^8$ non-uniform knots. For this computation, we adopt the weight distribution used to solve the linear call option in Section~\ref{sec: Linear Black Scholes}. The result is shown in Figure~\ref{fig:Leland_surface_slice_using_Weighted_nonuniform_NURBS} (left). Comparison with the P1- and P2-FEM solution is presented in the right figure. For P1-FEM and P2-FEM, $2^9$ and, respectively, $2^8$ elements are used, resulting in $2^9$ basis functions. This comparison shows  a close match between the IGA and P1-FEM solution.

Finally, in Table~\ref{tab:Uprice1} we present the computed error of the solutions of the uniform, unweighted cubic NURBS IGA relative to the P1-FEM solution, at $t = 0$, for various number of knots, 
\begin{align}
  \varepsilon = \|  \hat{v}_{P1}(S,0) - \hat{v}_{NURBS}(S,0) \|_2.
\end{align}
As the number of knots increases, the misfit relative to the P1-FEM solutions decreases. The average error contraction factor is about $2.91$, suggesting a convergence rate of order $(\Delta x)^{1.5}$.

\begin{table}[h!]
\caption{Solution of the Leland model at $t = 0$, calculated with non-uniform, unweighted cubic NURBS. $r = 0.1$, $\sigma = 0.2$, $\hat{K} = \$100$, $Le \approx 0.8$ and $\Delta\tau/\Delta x^2 = 0.1$.} \label{tab:Uprice1}
\begin{center}
\begin{tabular}{cccc} \hline
$nE$ & $n_{\tau}$ &  $\varepsilon$ & Contraction factor \\ \hline \hline
$2^8$ & 80 & 0.451445 &  -  \\
$2^9$ & 320 &  0.157192 &  2.87\\
$2^{10}$ & 1280 &  0.050986 & 3.08 \\
$2^{11}$ & 5120 &  0.017970 &  2.83\\
$2^{12}$ & 20480 &  0.006319 & 2.84\\
\hline
\end{tabular}
\end{center}
\label{table:Error_estimates_Leland}
\end{table}


\subsection{The AFV model}

For the AFV model, we consider the case with parameters listed in Table~\ref{tab:parameter}, which are adopted from~\cite{Ayache2003}. The computational domain in the $S$-space corresponds to $\Omega = (-6,2)$ in the $x$-space (the transformed space). 

\begin{figure}[H]
\centering
\includegraphics[width=0.49\textwidth]{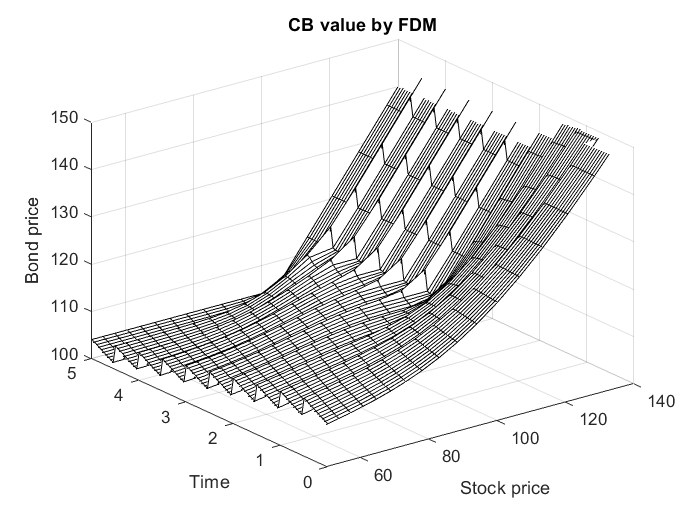}
\includegraphics[width=0.49\textwidth]{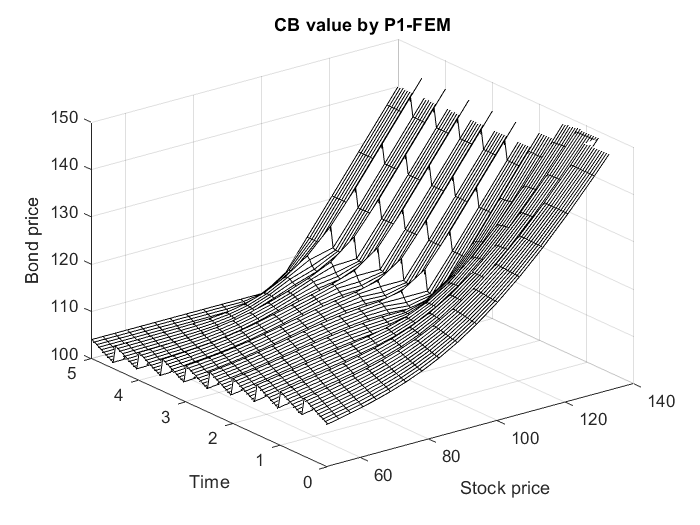}
\caption{Solution (bond price, $U$) of the AFV model at $t = 0$, using parameters in Table~\ref{tab:parameter}, $n_E = 2^7$ and $n_{\tau}=100$. Left figure: FDM; Right figure: P1-FEM.}
\label{fig:AFV_surf_FDM_P1}
\end{figure}

\begin{figure}[H]
\centering
\includegraphics[width=0.49\textwidth]{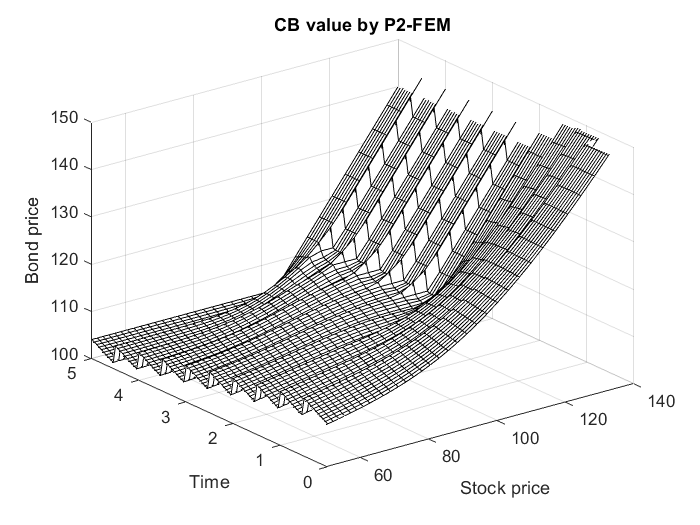}
\includegraphics[width=0.49\textwidth]{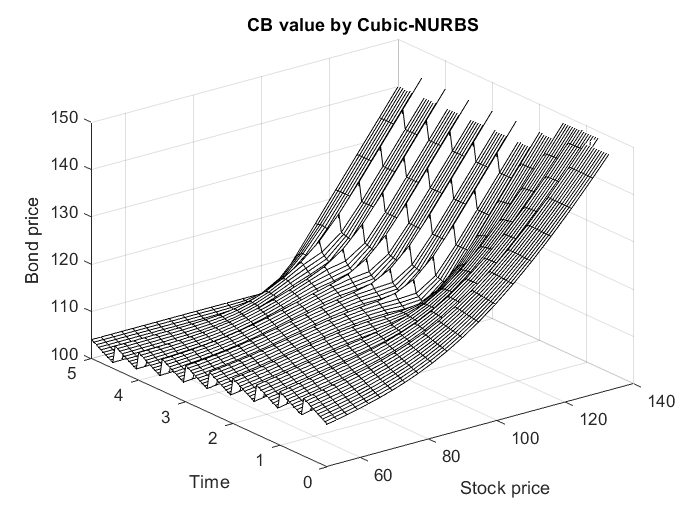}
\caption{Solution (bond price, $U$) of the AFV model at $t = 0$, using parameters in Table~\ref{tab:parameter}, $n_E = 2^7$ and $n_{\tau}=100$. Left figure: P2-FEM; Right figure:   unweighted cubic NURBS IGA with uniform knots.}
\label{fig:AFV_surf_P2_C2}
\end{figure}

Figures~\ref{fig:AFV_surf_FDM_P1}--\ref{fig:AFV_surf_P2_C2} show the bond price $U$ computed using FDM, P1-FEM, P2-FEM, and unweighted cubic NURBS on uniform mesh/knots, with the number of elements $nE = 2^7$ ($\Delta x = 0.0625$) and the number of time steps $n_{\tau} = 100$ ($\Delta \tau = 0.05$), corresponding to $\Delta \tau/\Delta x = 0.8$ and $\Delta \tau/\Delta x^2 = 12.8$. 
Comparing the solution of these methods, we hardly observe any noticeable difference in the surface of the solutions. We note that the jumps in the solution surfaces along the time axis take place at the time where coupons are paid. Even though $\Delta \tau/\Delta x^2 > 1$, no sign of instability is observable in the solutions.

\begin{figure}[H]
\centering
\includegraphics[width=0.49\textwidth]{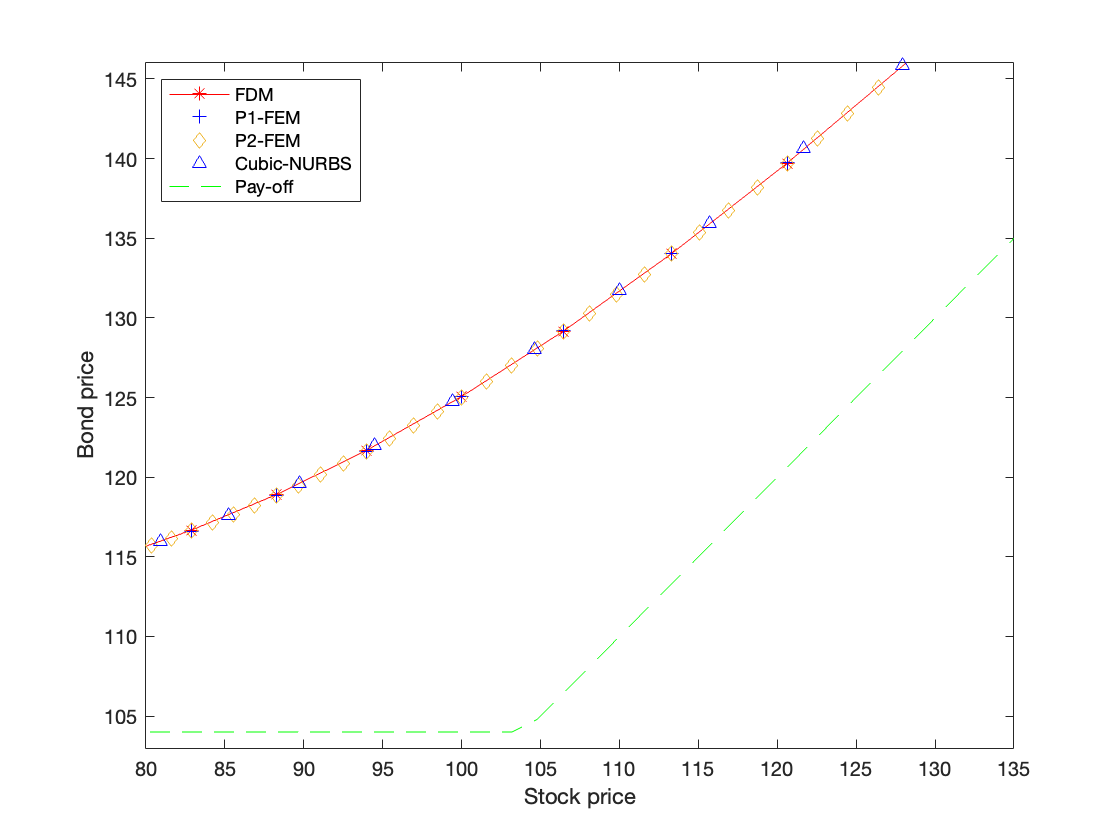}
\includegraphics[width=0.49\textwidth]{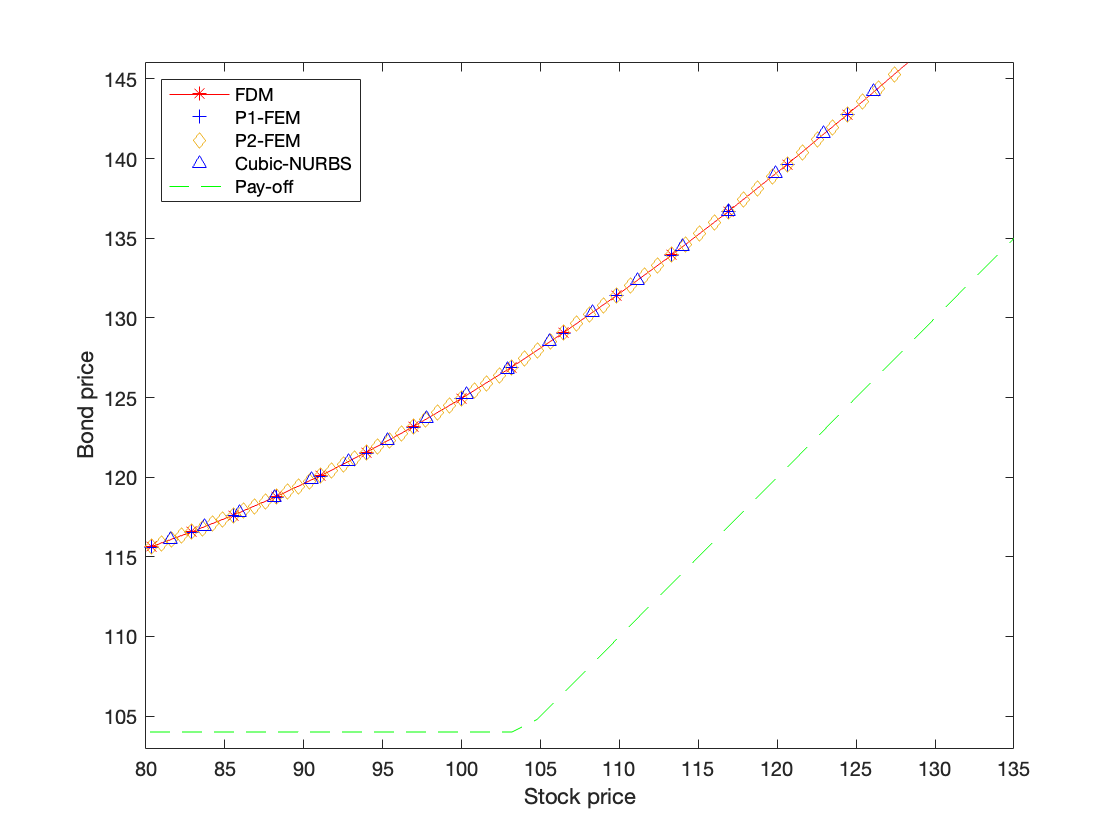}
\caption{Solution of the AFV model $t = 0$, $r = 0.05$, $\sigma = 0.2$, $F = \$100$, $\rho = 10^6$, $\eta = 0$. Left figure: $nE = 2^8$, $n_{\tau} = 200$; Right figure:  $nE = 2^9$, $n_{\tau} = 400$.}
\label{fig:AFV_2D_all}
\end{figure}

Solutions at $t = 0$ are shown in Figure~\ref{fig:AFV_2D_all} for $nE = 2^8$ and $n_{\tau} = 200$, and for $nE = 2^9$ and $n_{\tau} = 400$; both results in $\Delta \tau/\Delta x = 0.8$ and $\Delta \tau/\Delta x^2 > 1$. From the plot, the solution of FDM, P1-FEM, P2-FEM, and various NURBS-based FEM practically coincide, with the convertible bond price $U \approx 124.8745$ at $S = 100$. Numerical instabilities are also not detected in this case.



\begin{table}[!h]
\caption{Modeling and computational parameters.} \label{tab:parameter}
\begin{center}
\begin{tabular}{ |p{6.5cm}||p{6.5cm}|  }
 \hline
 \textbf{Parameter} & \textbf{Value}\\
 \hline
Time to maturity $T$ & 5 years \\
Conversion & 0 to 5 years into $k$ shares \\
Conversion ratio \textit{$k$} & 1.0 \\
Face Value $F$ & \$100 \\
Clean call price $B_{call}$ & \$110, from Year 3 to Year 5 ($t \in (2,5]$) \\
Clean put price $B_{put}$ & \$105, during Year 3 ($t = 3$)\\
Coupon payments $K_{coup}$ & \$4.0\\
Coupon dates & .5, 1.0, 1.5, ... ,5.0 (semiannual)\\
Risk-free rate $r$ & 0.05\\
Recovery factor $\hat{R}$ & 0\\
Hazard rate $p$ & 0.02\\
Default parameter $\eta$& 0 (Partial default )\\
Volatility $\sigma$ & 20\% or 0.2\\
Underlying stock price at $t = 0$, $S_{int}$  & \$100\\
Penalty parameter $\rho$ & $10^{6}$\\
Newton-Raphson's method tolerance  $tol$ & $10^{-6}$\\
\hline
\end{tabular}
\end{center}
\end{table}

Table~\ref{tab:AFV_table_non_uniform_new} summarizes the convergence of the unweighted cubic NURBS IGA to the solution at  $t = 0$ and $S = F = 100$. For a uniform mesh/knots, we vary $nE$ and $n_\tau$ such that the ratio $\Delta \tau/\Delta x$ remains constant at 0.8. As the element size and the time step decrease to 0, the solution of the unweighted cubic NURBS IGA converges to 124.87 (accurate to 2 decimal places), with a rate comparable to that of FDM, P1-FEM, and P2-FEM (shown also in the table). Given that the number of knots in the cubic NURBS is approximately equal to the number of basis functions in P1-FEM or unknowns in FDM, unweighted cubic NURBS IGA does not offer significant computational advantages over FDM or P1-FEM. Additionally, we computed the solutions using non-uniform, unweighted cubic NURBS, which are presented in the last column of Table~\ref{tab:AFV_table_non_uniform_new}. While these solutions also converge to 124.87, there is no noticeable improvement in the convergence rate compared to the uniform case, even as the number of knots increases.

\begin{table}[h!]
\caption{Convertible bond price based on the AFV model at $t = 0$ ans $S = 100$, computed using unweighted cubic NURBS, FDM, P1-FEM, and P2-FEM. The numerical parameters are given in Table~\ref{tab:parameter}.}  \label{tab:AFV_table_non_uniform_new}
\begin{center}
\begin{tabular}{c|cccc|cc} \hline 
& & & & & \multicolumn{2}{c}{\underline{Unweighted NURBS}} \\ 
$nE$ &  $n_\tau$ & FDM & P1-FEM & P2-FEM &  Uniform & Nonuniform  \\ \hline \hline
$2^{6}$&50&125.1718&125.0198&124.6675&125.5205&125.2426 \\ 
$2^{7}$ &100 &  125.0613 & 125.0557 & 125.0045&  125.1139 & 125.0675 \\
$2^{8}$ &200 &  124.9504 & 124.9210 & 124.9485 & 124.9676 & 124.9579 \\
$2^{9}$ &400 &  124.9123 & 124.9000 & 124.9094 & 124.9154 & 124.9115  \\
$2^{10}$ & 800 &  124.8914 & 124.8867 & 124.8893 & 124.8895 & 124.8898   \\
$2^{11}$ &1600 &   124.8805 & 124.8786 & 124.8789 & 124.8798 & 124.8795  \\
$2^{12}$ &3200 &   124.8749 & 124.8739 & 124.8745 & 124.8746 & 124.8745    \\
\hline
\end{tabular}
\end{center}
\end{table}

In Table~\ref{tab:AFV_table_non_uniform_optimal}, we present solutions of non-uniform, weighted cubic NURBS IGA. The weight distribution referred to as "optimal" produces solutions that match those of FDM and P1-FEM to two decimal places. This "optimal" distribution maintains the same functional form as in the European call case but differs in scale.  The ``optimal'' distribution retains the same shape as in the European call case, but with different scales. For non-optimal cases, we examine two scenarios: (i) using the optimal weights from the linear Black-Scholes problem in Section~\ref{sec: Linear Black Scholes}, and (ii) employing adjusted weights derived for the linear Black-Scholes problem. Both scenarios yield solutions that deviate from those of FDM and P1-FEM, within 2 decimal places. As the number of knots increases, the solutions exhibit convergence but generally approach values different from those reported in Table~\ref{tab:AFV_table_non_uniform_new}. These results highlight the sensitivity of IGA NURBS convergence to the choice of weights in the cubic NURBS basis functions. While constructing an optimal weight distribution is feasible, as demonstrated in the first column of the table, an automated method for selecting appropriate weights would enhance the generality and practicality of the approach. One possible strategy involves employing nonlinear optimization~\cite{Wall2008,laurent1993optimization}, which we propose to explore in future work.



\begin{table}[h!]
\caption{Convertible bond price based on the AFV model at $t = 0$ ans $S = 100$, computed using non-uniform, weighted cubic NURBS. The numerical parameters are given in Table~\ref{tab:parameter}.}  \label{tab:AFV_table_non_uniform_optimal}
\begin{center}
\begin{tabular}{c|c|ccc} \hline 
& & \multicolumn{3}{c}{ \underline{Weights}} \\ 
$nE$ &  $n_\tau$ & Optimal & Adjusted & Non-optimal  \\ \hline \hline
$2^{6}$ & 50 &  124.8745 & 124.8777  &  125.1409 \\ 
$2^{7}$ & 100 & 124.8745 & 124.6833  & 124.9433  \\
$2^{8}$ & 200 & 124.8745 & 124.5956  & 124.8542  \\
$2^{9}$ & 400 &          & 124.5541  & 124.8120   \\
$2^{10}$ & 800 &         & 124.5339  & 124.7915   \\
\hline
\end{tabular}
\end{center}
\end{table}


\begin{figure}[H]
\centering
\includegraphics[width=0.49\textwidth]{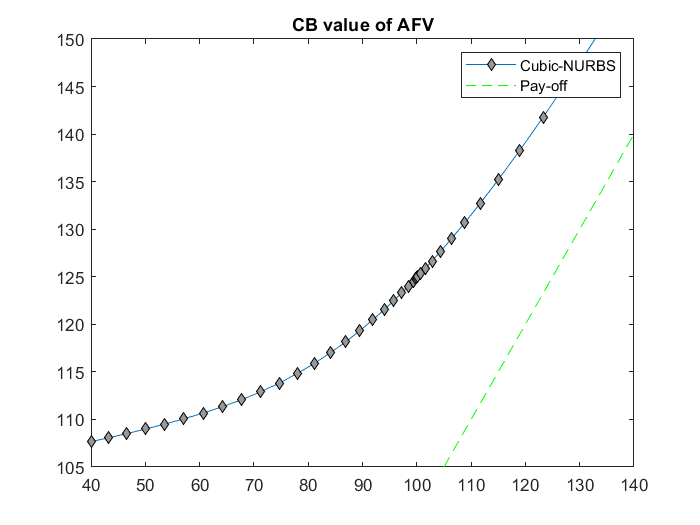}
\includegraphics[width=0.49\textwidth]{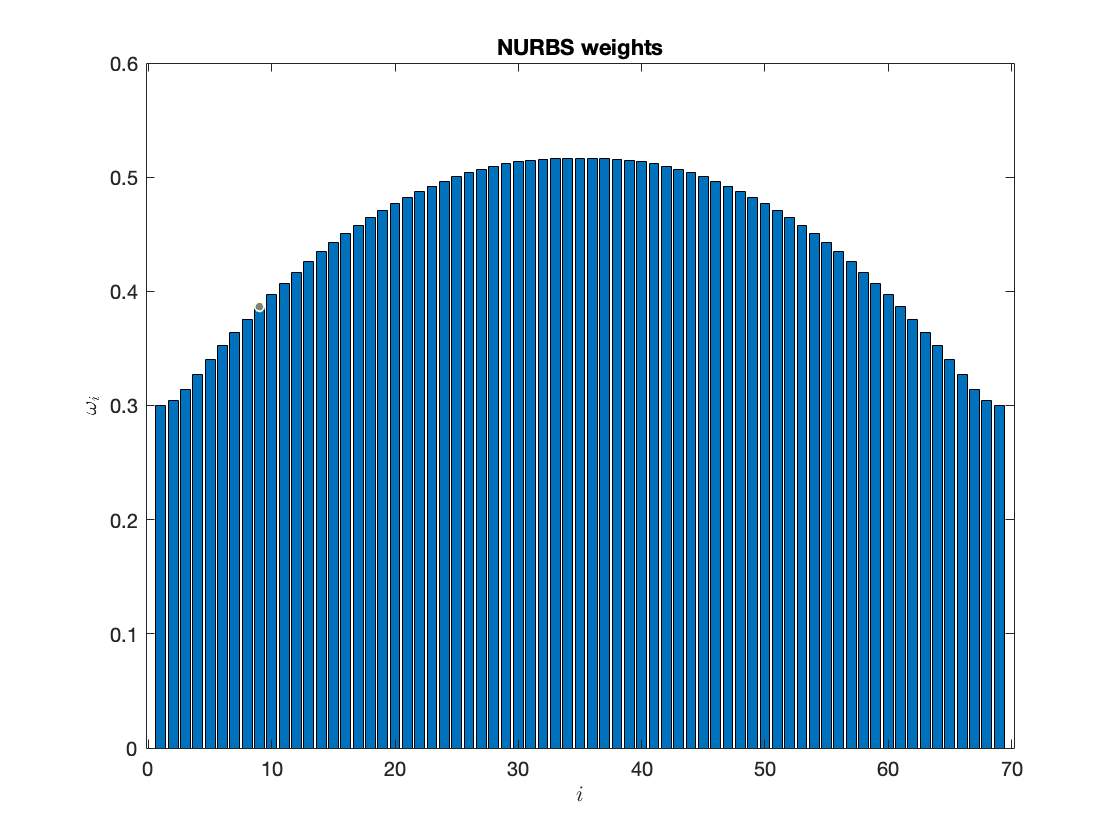}
\caption{Solution of the AFV model $t = 0$, $r = 0.05$, $\sigma = 0.2$, $F = \$100$, $\eta = 0$. Left figure: Non-uniform meshing with $nE = 2^6$, $n_{\tau} = 50$; Right figure: NURBS weights.}
\label{fig:AFV_nonuniform_fitted_new2}
\end{figure}

\subsection{Computation of Greeks}
For practitioners, the knowledge of the Greek values, such as Delta ($\Delta$), Gamma ($\Gamma$), and Theta ($\Theta$), is beneficial for developing hedging and risk management scenarios \cite{desmond, hull2003options, wilmott}. For instance, as the derivative of the solution, $\Delta$  indicates the tendency the convertible bond price $U$ to change as the underlying asset increases. As the second derivative of the bond price, $\Gamma$ tracks the  change of $\Delta$ under the movement of the underlying asset. Furthermore, one can use $\Theta = \frac{\partial w}{\partial t}$, with $w \in \{\hat{v},U,B,C\}$, to observe the behavior of the convertible bond $U$ as time approaches maturity.

 A frequent problem  of the numerical computation of the Greeks, especially in nonlinear cases \cite{Forsyth2002,Yousuf2012}, is that the computed Greeks suffer from high-frequency oscillation. This oscillation is purely numerical artefacts. Because the computed Greeks become unreliable in this case, this phenomenon creates trouble for hedging and controlling the risk in reality.  
 
\begin{figure}[H]
\centering
\includegraphics[width=0.49\textwidth]{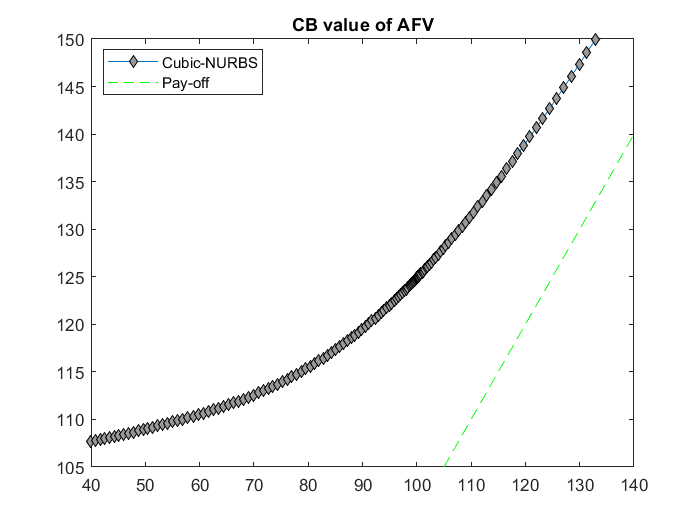}
\includegraphics[width=0.49\textwidth]{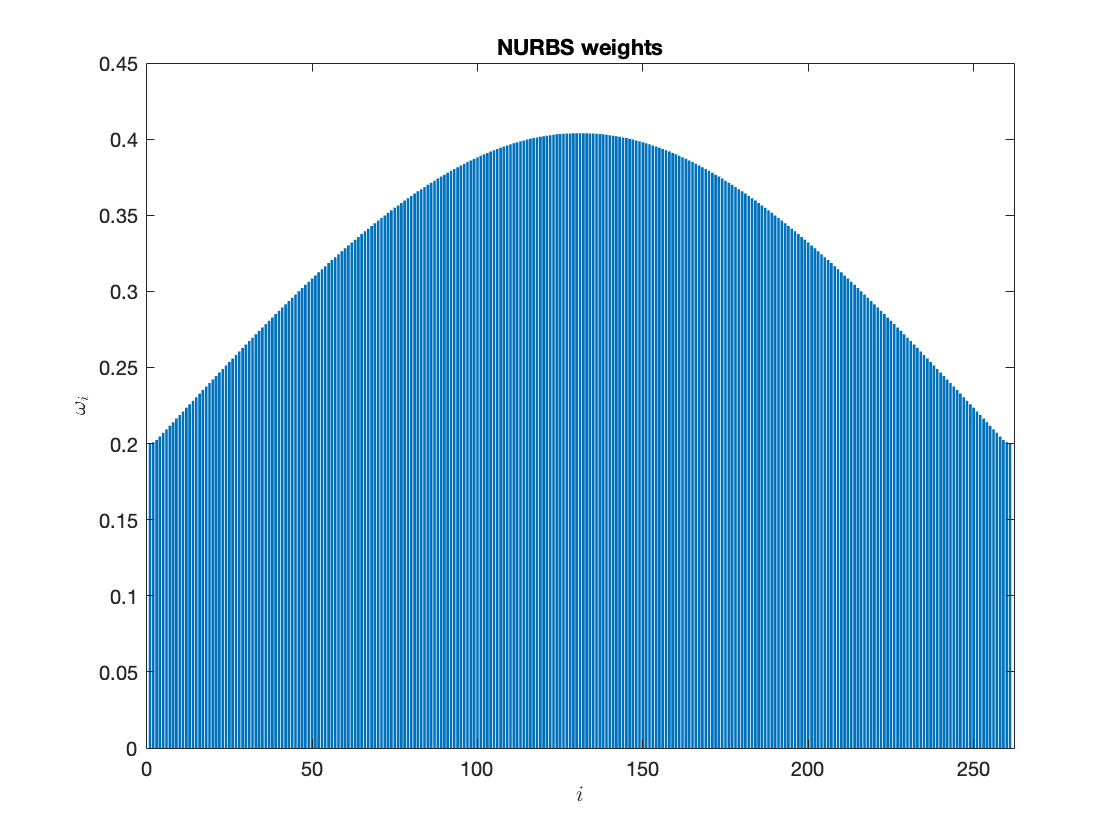}
\caption{Solution of the AFV model $t = 0$, $r = 0.05$, $\sigma = 0.2$, $F = \$100$, $\eta = 0$. Left figure: Non-uniform meshing with $nE = 2^8$, $n_{\tau} = 100$; Right figure: NURBS weights.}
\label{fig:AFV_nonuniform_fitted}
\end{figure}

The most common practice for the computation of the Greeks is by using finite differences based on the numerically computed option price. This approach can in principle be applied to solutions computed by IGA-NURBS. For the latter, since the solution itself is expressible as a linear combination of twice-differentiable basis functions (in the case of cubic NURBS), some Greeks, such as $\Delta$ and $\Gamma$, can however be computed via differentiation of the solution. Considering the solution of the form of the linear combinationx~\eqref{eq:lin.comb. of w}, these Greeks can in turn be expressed as a linear combination of derivatives of the basis functions. For example,
\begin{align}
    \frac{\partial w}{\partial x} =\sum_j w_j \frac{\partial R_{j,p}(x)}{\partial x} = \frac{|\Omega_{\xi}|}{|\Omega|} \sum_j w_j  \frac{\partial R_{j,p}(\xi)}{\partial \xi}. \label{eq:dwdx}
\end{align}
Considering $\Delta$ for the Leland's model, by transforming back the solution to the physical ($S$) space, we have
\begin{align}
    \Delta = \frac{\partial \hat{v}}{\partial S} = \frac{1}{S}e^{-\frac{1}{2}k\sigma^2(T-t)} \frac{|\Omega_{\xi}|}{|\Omega|}\sum_j w_j  \frac{\partial R_{j,p}(\xi)}{\partial \xi}.
\end{align}
$\Gamma$ can also be computed in the similar fashion, where, after transforming back to the $S$-space, is given by
\begin{align}
    \Gamma = \frac{\partial^2\hat{v}}{\partial S^2} = \frac{1}{S^2}e^{-\frac{1}{2}k\sigma^2(T-t)} \frac{|\Omega_{\xi}|}{|\Omega|}\left( \frac{|\Omega_{\xi}|}{|\Omega|}\sum_j w_j  \frac{\partial^2 R_{j,p}(\xi)}{\partial \xi} - \sum_j w_j \frac{\partial R_{j,p}(\xi)}{\partial \xi}\right).
\end{align}
As the time derivative is approximated via finite differences -- Crank-Nicolson in our case -- it is more natural to approximate $\Theta$ by a finite difference scheme and compute this approximation using the IGA-NURBS solution.

\begin{figure}[H]
\centering
\includegraphics[width=0.4\textwidth]{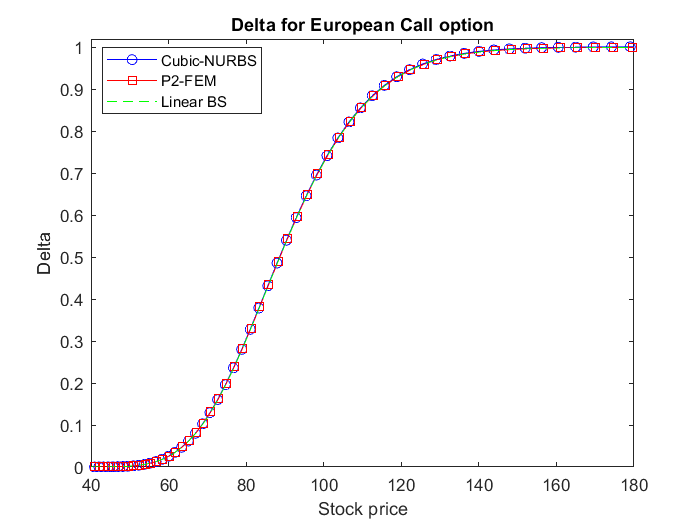}
\includegraphics[width=0.4\textwidth]{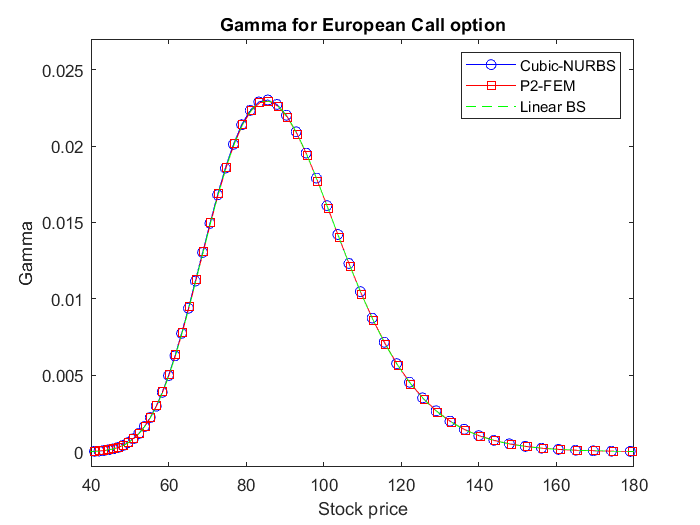}
\includegraphics[width=0.4\textwidth]{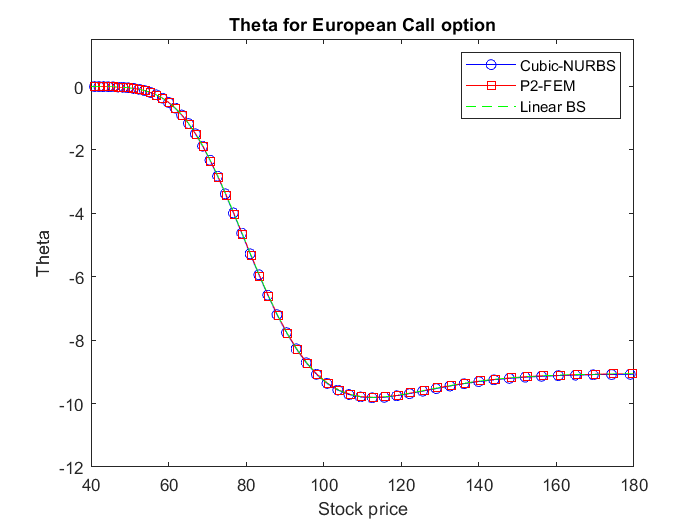}
\caption{Greeks for European call option with the parameters $t = 0$, $r = 0.1$, $r_c = 0.02$, $\sigma = 0.2$, $\hat{K} = \$100$. For Cubic-NURBS:  $nE = 2^{12}$, $n_{\tau} = 2\times 10^4$; For P2-FEM: $nE = 2^{11}$, $n_{\tau} = 2\times 10^4$;  Top left: Delta; Top right: Gamma;  Bottom: Theta.}
\label{fig:greeks_exact_nurbs_P2}
\end{figure}

In Figure~\ref{fig:greeks_exact_nurbs_P2}, we present the values of  $\Delta$, $\Gamma$, and $\Theta$ for the linear European call option ($Le = 0$) numerically computed  using Cubic-NURBS at $t = 0$. We compare them with the exact Greeks and those derived from the P2-FEM solution. In this linear case, both the Cubic-NURBS and P2-FEM~\cite{kazbek2024phdthesis} methods yield Greeks that closely approximate the exact values.

\begin{figure}[H]
\centering
\includegraphics[width=0.32\textwidth]{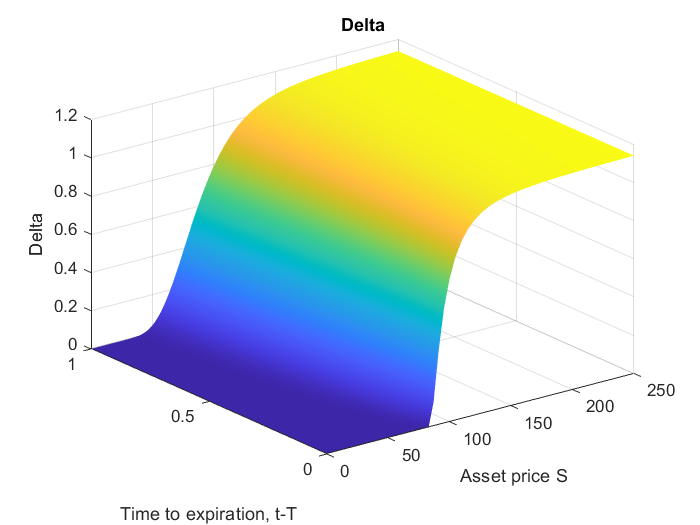}
\includegraphics[width=0.32\textwidth]{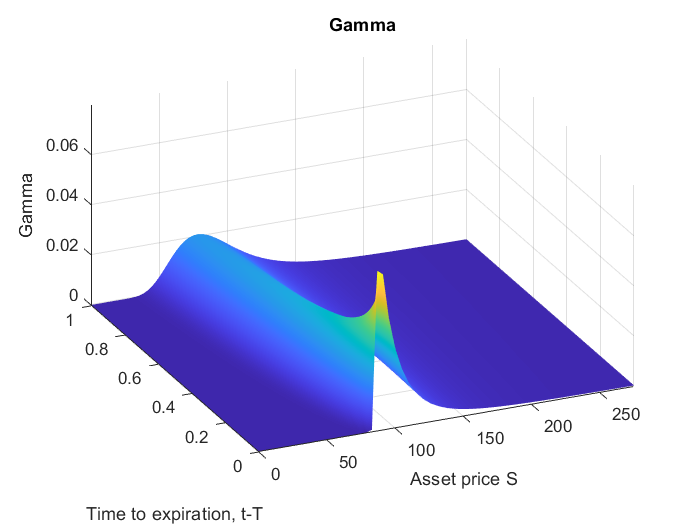}
\includegraphics[width=0.32\textwidth]{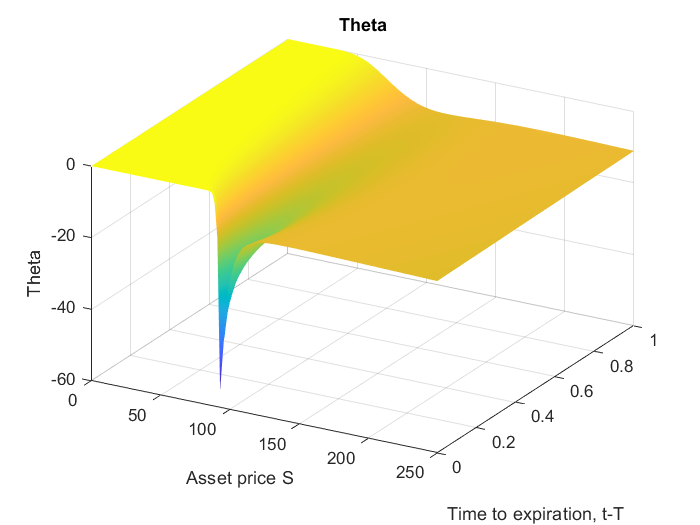}
\caption{Greeks for Leland model with $t = 0$, $r = 0.1$, $\sigma = 0.2$, $\hat{K} = \$100$, $Le \approx 0.8$, $\Delta x = 0.0017$, $\Delta\tau = 7\times 10^{-7}$, therefore, $\Delta\tau/\Delta x = 4\times 10^{-4}$, $\Delta\tau/\Delta x^2 = 0.2$. Left: Delta; Middle: Gamma; Right: Theta}
\label{fig:greeks_surf_Leland_nurbs}
\end{figure}

\begin{figure}[H]
\centering
\includegraphics[width=0.37\textwidth]{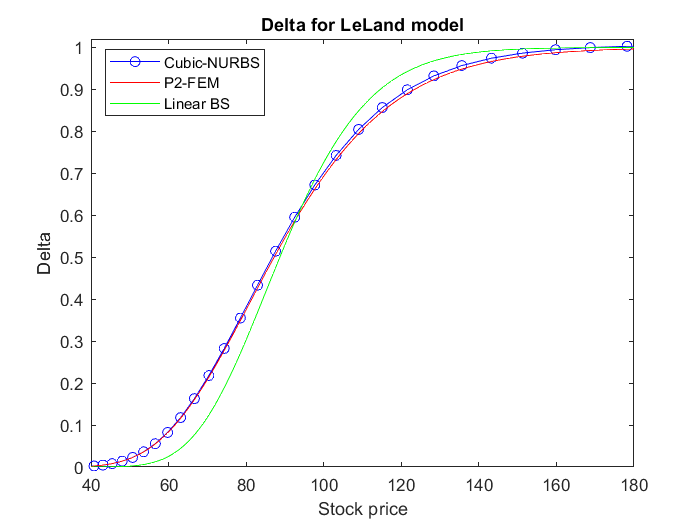}
\includegraphics[width=0.37\textwidth]{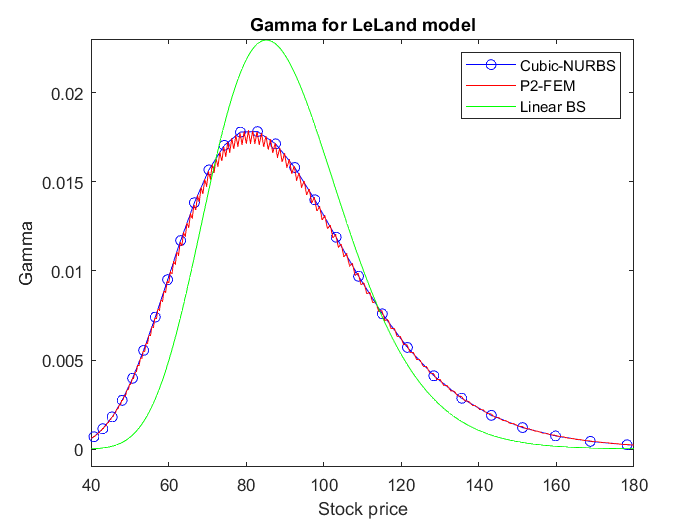}
\includegraphics[width=0.37\textwidth]{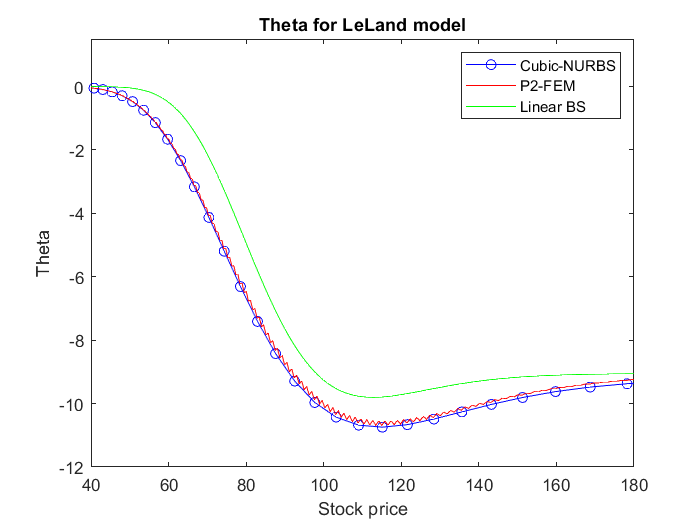}
\caption{Greeks for Leland model with $t = 0$, $r = 0.1$, $\sigma = 0.2$, $\hat{K} = \$100$, $Le \approx 0.8$. For NURBS: $\Delta x = 0.005$, $\Delta\tau = 10^{-6}$, $\Delta\tau/\Delta x = 1\times 10^{-4}$, $\Delta\tau/\Delta x^2 = 0.02$;
For P2-FEM: $\Delta x = 0.013$, $\Delta\tau = 5\times 10^{-7}$, $\Delta\tau/\Delta x = 3.6\times 10^{-5}$, $\Delta\tau/\Delta x^2 = 0.02$.  Top left: Delta; Top right: Gamma;  Bottom: Theta.}
\label{fig:greeks_2D_Leland_nurbs}
\end{figure}

Greeks for the nonlinear Leland model with $Le = 0.8$ are shown in Figure~\ref{fig:greeks_surf_Leland_nurbs}, computed using the cubic-NURBS. Apart along the terminal time $t = T$, where $\Gamma$ and $\Theta$ are discontinuous at $S = \hat{K} = 100$ (the strike price), the surface of the Greeks are smooth. In Figure~\ref{fig:greeks_2D_Leland_nurbs}, we show the slice of the Greeks surface at the initial time ($t=0$). We compare the Greeks computed using IGA-NURBS with those from P2-FEM. The Greeks from the linear case are shown to illustrate the effects of transaction costs on the Greeks in the modeling. Notably, $\Delta$
computed with cubic-NURBS aligns well with P2-FEM results, while $\Gamma$ and $\Theta$ exhibit significantly different behaviors. The P2-FEM solutions produce non-smooth, oscillatory 
$\Gamma$ and $\Theta$, particularly pronounced near the strike price ($S = \hat{K} = 100$. In contrast, the values obtained from cubic NURBS are smooth and free of oscillations.  This suggests that the Greeks computed using cubic NURBS may be more reliable than those from P2-FEM.

\begin{figure}[H]
\centering
\includegraphics[width=0.37\textwidth]{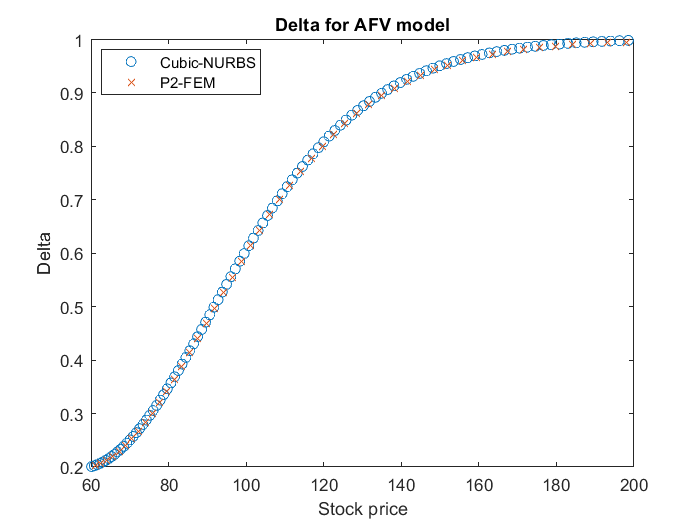}
\includegraphics[width=0.37\textwidth]{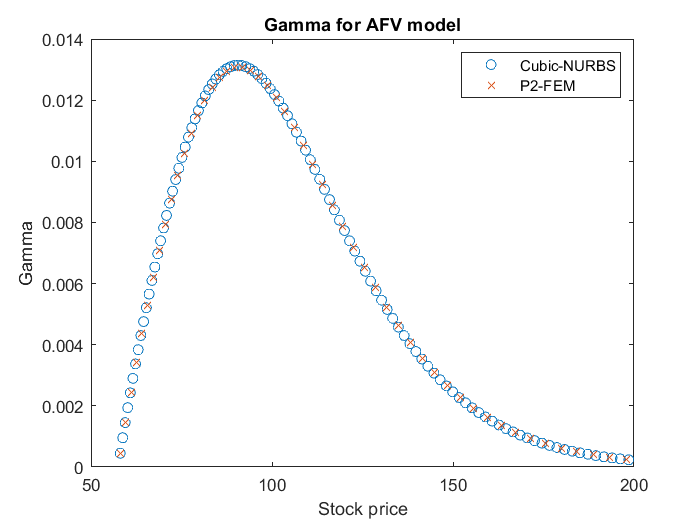}
\includegraphics[width=0.37\textwidth]{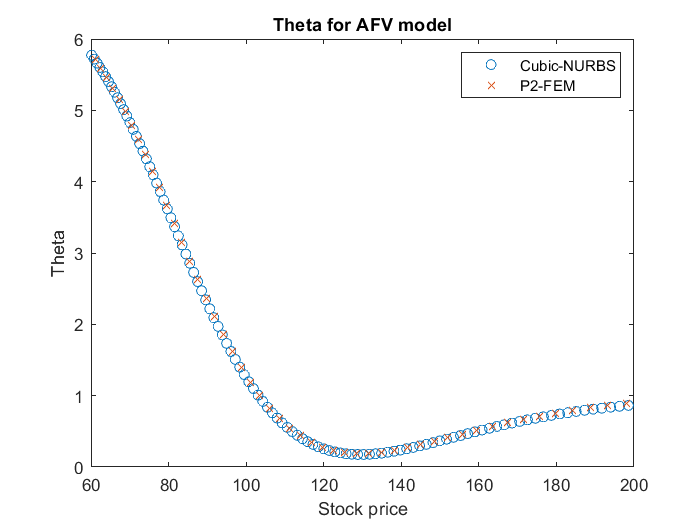}
\caption{Greeks for AFV model with $t = 0$, $r = 0.05$, $\sigma = 0.2$, $F = \$100$. For NURBS: $nE = 2048$, $n_{\tau} = 2000$;
For P2-FEM: $nE = 1000$, $n_{\tau} = 2000$.  Top left: Delta; Top right: Gamma;  Bottom: Theta.}
\label{fig:greeks_for_AFV}
\end{figure}

Greeks related to the convertible bond values $U$ can be also be computed based on~\eqref{eq:dwdx} and the appropriate transformation back to the $S$-space. In this case, since the solution $U$ at the time of coupon payment is not differentiable, the Greeks can not be smooth everywhere. We present the Greeks of the AFV model at the initial time $t = 0$ in~\ref{fig:greeks_for_AFV}, based on the cubic-NURBS and P2-FEM. In this case, the two methods yield Greeks which align well and are smooth.

\noindent

\section{Conclusion}
\label{sec:conclusion}
In this paper, we discussed IGA for solving nonlinear option pricing. We considered two models: the European contract with transaction costs modeled by the Leland PDE and the nonlinear American contract for convertible bond pricing modeled by the AFV model. We compared the numerical results of the cubic NURBS with FDM and P1/P2-FEM and observed that solutions of the cubic NURBS align well with those of FDM and P1-FEM. Furthermore, this solution could be obtained by the cubic NURBS IGA  using relatively small number of knots (or basis functions), when appropriately chosen weights were used, reducing the computational time significantly. As the cubic NURBS basis functions are $C^2$-continuous, the calculations of Greeks ($\Delta$ and $\Gamma$) could be done in a natural way using the the derivatives of the basis functions.
\section*{Declaration}
Everyone who contributed to this work was aware of the stage of the manuscript and agreed to submit it to the journal without any conflict of interests. All data and information used in this work are properly cited.
\section*{Data availability}
MATLAB P-codes for benchmarking and validation are available from the corresponding author upon reasonable request.
\section*{Funding}
This research was conducted without any financial support from external funding agencies, institutions, or organizations.


  

 
\bibliographystyle{style} 
\newcommand{\noopsort}[1]{} \newcommand{\printfirst}[2]{#1} \newcommand{\singleletter}[1]{#1} \newcommand{\switchargs}[2]{#2#1}

\end{document}